\documentclass[10pt,final,twocolumn]{IEEEtran}
\usepackage{amsmath,amssymb,amsthm,epsfig,dsfont,color,empheq,enumerate}
\usepackage{algpseudocode,algorithm}
\usepackage{graphicx,subfigure}
\usepackage[usenames,dvipsnames]{xcolor}
\usepackage{stfloats,url}
\usepackage{bbm}

\DeclareMathOperator{\sign}{sign}

\newtheorem{proposition}{Proposition}

\theoremstyle{remark}

\newtheoremstyle{remboldstyle}
{}{}{\itshape}{}{\bfseries}{.}{.5em}{{\thmname{#1 }}{\thmnumber{#2}}{\thmnote{ (#3)}}}
\theoremstyle{remboldstyle}
\newtheorem{rembold}{Remark}

\begin{document}
\title{Online Censoring for Large-Scale Regressions\\ 
	with Application to Streaming Big Data}

\author{Dimitris Berberidis,~\IEEEmembership{Student Member,~IEEE,}
Vassilis Kekatos,~\IEEEmembership{Member,~IEEE,}
and\\ Georgios B. Giannakis*,~\IEEEmembership{Fellow,~IEEE}%
\thanks{Work in this paper was supported by the Institute of Renewable Energy and
the Environment grant no. RL-0010-13, University of Minnesota,
and NSF grants 1343860, 1442686, and 1500713. }
\thanks{D. Berberidis, V. Kekatos, and G. B. Giannakis are with the ECE Dept., University of Minnesota, Minneapolis, MN 55455, USA. E-mails:\{bermp001,kekatos,georgios\}@umn.edu.}
}

\maketitle
\begin{abstract}
Linear regression is arguably the most prominent among statistical inference methods, popular both for its simplicity as well as its broad applicability. On par with data-intensive applications, the sheer size of linear regression problems creates an ever growing demand for quick and cost efficient solvers. Fortunately, a significant percentage of the data accrued can be omitted while maintaining a certain quality of statistical inference with an affordable computational budget.  The present paper introduces means of identifying and omitting ``less informative'' observations in an online and data-adaptive fashion, built on principles of stochastic approximation and data censoring. First- and second-order stochastic approximation maximum likelihood-based algorithms for censored observations are developed for estimating the regression coefficients. Online algorithms are also put forth to reduce the overall complexity by adaptively performing censoring along with estimation. The novel algorithms entail simple closed-form updates, and have provable (non)asymptotic convergence guarantees. Furthermore, specific rules are investigated for tuning to desired censoring patterns and levels of dimensionality reduction. Simulated tests on real and synthetic datasets corroborate the efficacy of the proposed data-adaptive methods compared to data-agnostic random projection-based alternatives.
\end{abstract}

\section{Introduction}
Nowadays omni-present monitoring sensors, search engines, rating sites, and Internet-friendly portable devices generate massive volumes of typically dynamic data~\cite{bigdata}. The task of extracting the most informative, yet low-dimensional structure from high-dimensional datasets is thus of utmost importance. Fast-streaming and large in volume data, motivate well updating analytics rather than re-calculating new ones from scratch, each time a new observation becomes available. Redundancy is an attribute of massive datasets encountered in various applications~\cite{cs}, and exploiting it judiciously offers an effective means of reducing data processing costs.

In this regard, the notion of optimal design of experiments has been advocated for reducing the number of data required for inference tasks~\cite{pukelsheim1993optimal}. In recent works, the importance of sequential optimization along with random sampling of Big Data has been highlighted~\cite{bigdata}. Specifically for linear regressions, random projection (RP)-based methods have been advocated for reducing the size of large-scale least-squares (LS) problems~\cite{dmmm06acm,bd09laa,mahoney2011ftml}. As far as online alternatives, the randomized Kaczmarz's (a.k.a. normalized least-mean-squares (LMS)) algorithm generates a sequence of linear regression estimates from projections onto convex subsets of the data~\cite{strohmer2009randomized,sgdrka,agaskar2014randomized}. Sequential optimization includes stochastic approximation, along with recent advances on online learning~\cite{shalev2011online}. Frugal solvers of (possibly sparse) linear regressions are available by estimating  regression coefficients based on (severely) quantized data \cite{alej_quant,yaniv_1bit}; see also \cite{gonz_dist} for decentralized sparse LS solvers.

In this context, the idea here draws on interval censoring to discard ``less informative'' observations. Censoring emerges naturally in several areas, and \emph{batch} estimators relying on censored data have been used in econometrics, biometrics, and engineering tasks \cite{tobit}, including survival analysis~\cite{bio2008evers}, saturated metering~\cite{tobit1}, and spectrum sensing~\cite{ml13jsac}. It has recently been employed to select data for distributed estimation of parameters and dynamical processes using resource-constrained wireless sensor networks, thus trading off performance for tractability~\cite{tsp2012ggeric,tsp2013youxie,arxiv2014wang}. These works confirm that estimation accuracy achieved with censored measurements can be comparable to that based on uncensored data. Hence, censoring offers the potential to lower data processing costs, a feature certainly desirable in Big Data applications.

To this end, the present work employs interval censoring for large-scale \emph{online} regressions. Its key novelty is to sequentially test and update regression estimates using censored data. Two censoring strategies are put forth, each tailored for mitigating different costs. In the first one, stochastic approximation algorithms are developed for sequentially updating the regression coefficients with low-complexity first- or second-order iterations to maximize the likelihood of censored and uncensored observations. This strategy is ideal when the number of observations are to be reduced, in order to lower the cost of storage or transmission to a remote estimation site. Relative to~\cite{tsp2012ggeric,tsp2013youxie}, the contribution here is a novel online scheme that greatly reduces storage requirements without requiring feedback from the estimator to sensors. Error bounds are derived, while simulations demonstrate performance close to estimation error limits.

The second censoring strategy focuses on reducing the complexity of large-scale linear regressions. The proposed methods are also online by design, but may also be readily used to reduce the complexity of solving a batch linear regression problem. The difference with dimensionality-reducing alternatives, such as optimal design of experiments, randomized Kaczmarz's and  RP-based methods, is that the introduced technique reduces complexity in a data-driven manner. 

The rest of the paper is as follows. A formal problem description is in Section~\ref{sec:problem}, while the two censoring rules are introduced in Section~\ref{subsec:ACNACrule}. First- and second-order stochastic approximation maximum-likelihood-based algorithms for censored observations are developed in Section~\ref{sec:NAConline}, along with threshold selection rules for controlled data reduction in Section~\ref{subsec:NACprob}. Adaptive censoring algorithms for reduced-complexity linear regressions are in Section~\ref{sec:ACstreaming}, with corresponding threshold selection rules given in Section~\ref{subsec:ACcontrol}, and robust versions of the algorithms outlined in Section~\ref{subsec:robust}. The proposed online-censoring and reduced-complexity methods are tested on synthetic as well as real data, and compared with competing alternatives in Section~\ref{sec:Sim}. Finally, concluding remarks are made in Section~\ref{sec:conclusion}.   

\emph{Notation.} Lower- (upper-) case boldface letters denote column vectors (matrices). Calligraphic symbols are reserved for sets, while symbol $^T$ stands for transposition. Vectors $\mathbf{0}$, $\mathbf{1}$, and $\mathbf{e}_n$ denote the all-zeros, the all-ones, and the $n$-th canonical vector, respectively. Notation $\mathcal{N}(\mathbf{m},\mathbf{C})$ stands for the multivariate Gaussian distribution with mean $\mathbf{m}$ and covariance matrix $\mathbf{C}$. The $\ell_1$- and $\ell_2$-norms of a vector $\mathbf{y}\in\mathbb{R}^d$ are defined as $\|\mathbf{y}\|_1:=\sum_{i=1}^d|y(i)|$ and $\|\mathbf{y}\|_2:=\sqrt{\sum_{i=1}^d|y(i)|^2}$, respectively; $\phi(t):=({1}/{\sqrt{2\pi}}){\mathrm{exp}}(-t^2/2)$ denotes the standardized Gaussian probability density function (pdf), and $Q(z):=\int_{z}^{+\infty}\phi(t){\mathrm{d}}t$ the associated complementary cumulative distribution function. Finally, for a matrix $\mathbf{X}$ let $\mathrm{tr(\mathbf{X})},~\lambda_{\min}(\mathbf{X})$ and $\lambda_{\max}(\mathbf{X})$ denote the trace, minimum and maximum eigenvalue, respectively. 

\section{Problem Statement and Preliminaries}\label{sec:problem}
Consider a $p\times 1$ vector of unknown parameters $\boldsymbol{\theta}_{o}$ generating scalar streaming observations 
\begin{equation}\label{model}
y_n=\mathbf{x}_n^T\boldsymbol{\theta}_o+\upsilon_n,~~~~n=1, 2, \ldots, D
\end{equation}
where $\mathbf{x}_n$ is the $n$-th row of the $D\times{p}$ regression matrix $\mathbf{X}$, and the noise samples $\upsilon_n$ are assumed independently drawn from $\mathcal{N}(0,\sigma^2)$. The high-level goal is to estimate $\boldsymbol{\theta}_{o}$ in an online fashion, while meeting minimal resource requirements. The term resources here refers to the total number of utilized observations and/or regression rows, as well as the overall computational complexity of the estimation task. Furthermore, the sought data-and complexity-reduction schemes are desired to be data-adaptive, and thus scalable to the size of any given dataset $\{y_n,\mathbf{x}_n\}_{n=1}^D$. To meet such requirements, the proposed first- and second-order online estimation algorithms are based on the following two distinct censoring methods.         

\subsection{NAC and AC Rules}\label{subsec:ACNACrule}
A generic censoring rule for the data in \eqref{model} is given by
\begin{equation}\label{crule}
z_n:=\left\{\begin{array}{ll}
\ast&,~y_n\in\mathcal{C}_n\\
y_n&,~{\text{otherwise}}
\end{array}\right.,~~~~n=1,\ldots,D
\end{equation}
where $\ast$ denotes an unknown value when the $n$-th datum has been censored (thus it is unavailable) - a case when we only know that $y_n\in\mathcal{C}_n$ for some set $\mathcal{C}_n$; otherwise, the actual measurement $y_n$ is observed. Given $\{z_n,\mathbf{x}_n\}_{n=1}^D$, the goal is to estimate $\boldsymbol{\theta}_o$. Aiming to reduce the cost of storage and possible transmission, it is prudent to rely on innovation-based interval censoring of $y_n$. To this end, define per time $n$ the binary censoring variable $c_n=1$ if $y_n\in\mathcal{C}_n$; and zero otherwise. Each datum is decided to be censored or not using a predictor $\hat{y}_n$ formed using a preliminary (e.g., LS) estimate of $\boldsymbol{\theta}_o$ as
\begin{equation}\label{LSE}
 \hat{\boldsymbol{\theta}}_{K}=(\mathbf{X}_K^T\mathbf{X}_K)^{-1}\mathbf{X}_K^T\mathbf{y}_K
\end{equation}
from $K\geq{p}$ measurements $(K\ll{D})$ collected in $\mathbf{y}_{K}$, and the corresponding $K\times{p}$ regression matrix $\mathbf{X}_K$. Given $\hat{y}_n =\mathbf{x}_n^T\hat{\boldsymbol{\theta}}_{K}$, the prediction error $\tilde{y}_n:=y_n-\hat{y}_n$ quantifies the importance of datum $n$ in estimating $\boldsymbol{\theta}_o$. The latter motivates what we term \emph{non-adaptive censoring} (NAC) strategy:
\begin{equation}\label{censoringrule}
(z_n,c_n):=\left\{\begin{array}{ll}
(y_n,0)&,~\mathrm{if}~\left|\frac{y_n-\mathbf{x}_n^T\hat{\boldsymbol{\theta}}_{K}}{\sigma}\right|\ge \tau_n\\
(\ast,1)&,~\mathrm{otherwise}
\end{array}
\right.
\end{equation}
where $\{\tau_n\}_{n=1}^D$ are censoring thresholds, and as in~\eqref{crule}, $*$ signifies that the exact value of $y_n$ is unavailable. The rule \eqref{censoringrule} censors measurements whose absolute normalized innovation is smaller than $\tau_n$; and it is non-adaptive in the sense that censoring depends on $\hat{\boldsymbol{\theta}}_K$ that has been derived from a fixed subset of $K$ measurements. Clearly, the selection of $\{\tau_n\}_{n=1}^{D}$ affects the proportion of censored data. Given streaming data $\{z_n,c_n,\mathbf{x}_n\}$, the next section will consider constructing a sequential estimator of $\boldsymbol{\theta}_o$ from censored measurements.

The efficiency of NAC in \eqref{censoringrule} in terms of selecting informative data depends on the initial estimate $\hat{\boldsymbol{\theta}}_{K}$. A data-adaptive alternative is to take into account all censored data $\{\mathbf{x}_i,z_i\}_{i=1}^{n-1}$ available up to time $n$. Predicting data through the most recent estimate $\hat{\boldsymbol{\theta}}_{n-1}$ defines our \emph{data-adaptive censoring} (AC) rule:
\begin{equation}\label{ACcensoringrule}
(z_n,c_n):=\left\{\begin{array}{ll}
(y_n,0)&,~\mathrm{if}~\left|\frac{y_n-\mathbf{x}_n^T\boldsymbol{\theta}_{n-1}}{\sigma}\right|\ge \tau_n\\
(\ast,1)&,~\mathrm{otherwise}
\end{array}\right..
\end{equation}
In Section~\ref{sec:ACstreaming}, \eqref{ACcensoringrule} will be combined with first- and second-order iterations to perform joint estimation and censoring online. Implementing the AC rule requires feeding back $\boldsymbol{\theta}_{n-1}$ from the estimator to the censor, a feature that may be undesirable in distributed estimation setups. Nonetheless, in centralized linear regression, AC is well motivated for reducing the problem dimension and computational complexity. 

\section{Online Estimation with NAC}\label{sec:NAConline}

Since noise samples $\{\upsilon_n\}_{n=1}^D$ in \eqref{model} are independent and \eqref{censoringrule} applies independently over data, $\{z_n, c_n\}_{n=1}^D$ are independent too. With $\mathbf{z}_D:=[z_1,\ldots,z_D]^T$ and $\mathbf{c}_D:=[c_1,\ldots,c_D]^T$, the joint pdf is $p(\mathbf{z}_D,\mathbf{c}_D;\boldsymbol{\theta})=\prod_{n=1}^D p(z_n,c_n;\boldsymbol{\theta})$ with
\begin{equation}\label{jointpdf}
p(z_n,c_n;\boldsymbol{\theta})=\left[\mathcal{N}\left(z_n;\mathbf{x}_n^T\boldsymbol{\theta},\sigma^2\right)\right]^{1-c_n}
\left[\Pr\{c_n=1\}\right]^{c_n}
\end{equation}
since $c_n=0$ means no censoring, and thus $z_n=y_n$ is Gaussian distributed; whereas $c_n=1$ implies $|y_n-\hat{y}_n|\leq{\tau_n\sigma}$, that is $\Pr\{c_n=1\}=\Pr\{\hat{y}_n-\tau_n\sigma-\mathbf{x}_n^T\boldsymbol{\theta}_0\leq{v_n}\leq{\hat{y}_n+\tau_n\sigma-\mathbf{x}_n^T\boldsymbol{\theta}_0}\}$, and after recalling that $v_n$ is Gaussian 
\begin{equation*}
\Pr\{c_n=1\}=Q\left(z_n^l(\boldsymbol{\theta})\right)-Q\left(z_n^u(\boldsymbol{\theta})\right)
\end{equation*}
where $z_n^l(\boldsymbol{\theta}):=-\tau_n-\frac{\mathbf{x}_n^T\boldsymbol{\theta}-\hat{y}_n}{\sigma}$ and $z_n^u(\boldsymbol{\theta}) := \tau_n-\frac{\mathbf{x}_n^T\boldsymbol{\theta}-\hat{y}_n}{\sigma}$. Then, the maximum-likelihood estimator (MLE) of $\boldsymbol{\theta}_o$ is
\begin{equation}\label{MLE}
\hat{\boldsymbol{\theta}}=\arg\min_{\boldsymbol{\theta}}~\mathcal{L}_D(\boldsymbol{\theta}):=\sum_{n=1}^D\ell_n(\boldsymbol{\theta})
\end{equation}
where functions $\ell_n(\boldsymbol{\theta})$ are given by (cf.~\eqref{jointpdf})
\begin{equation*}
\ell_n(\boldsymbol{\theta}):=\tfrac{1-c_n}{2\sigma^2}
\left(y_n-\mathbf{x}_n^T\boldsymbol{\theta}\right)^2 -c_n \log\left[Q\left(z_n^{l}(\boldsymbol{\theta})\right)-Q\left(z_n^u(\boldsymbol{\theta})\right)\right].
\end{equation*}
If the entire dataset $\{z_n,c_n,\mathbf{x}_n\}_{n=1}^D$ were available, the MLE could be obtained via gradient descent or Newton iterations.
 
Considering Big Data applications where storage resources are scarce, we resort to a stochastic approximation solution and process censored data sequentially. In particular, when datum $n$ becomes available, the unknown parameter is updated as
\begin{equation}\label{SGD}
\boldsymbol{\theta}_{n}:=\boldsymbol{\theta}_{n-1}-\mu_n\mathbf{g}_n(\boldsymbol{\theta}_{n-1}) 
\end{equation}
for a step size $\mu_n>0$, and with $\mathbf{g}_n(\boldsymbol{\theta}) =\beta_n(\boldsymbol{\theta})\mathbf{x}_n$ denoting the gradient of $\ell_n(\boldsymbol{\theta})$, where
\begin{equation}\label{beta}
\beta_n(\boldsymbol{\theta}):=\tfrac{1-c_n}{\sigma^2}(y_n-\mathbf{x}_n^T\boldsymbol{\theta})+\frac{c_n}{\sigma} \frac{\phi\left(z_n^u(\boldsymbol{\theta})\right)-\phi\left(z_n^l(\boldsymbol{\theta})\right)}{Q\left(z_n^u(\boldsymbol{\theta})\right)-Q\left(z_n^l(\boldsymbol{\theta})\right)}.
\end{equation}
The overall scheme is tabulated as Algorithm~\ref{SAMLEalg}.

\begin{algorithm}[t]
  \caption{Stochastic Approximation (SA)-MLE}
  \begin{algorithmic}
  \State Initialize $\boldsymbol{\theta}_0$ as the LSE $\hat{\boldsymbol{\theta}}_{K}$ in \eqref{LSE}.
  \For {$n=1:D$}
  \State Measurement $y_n$ is possibly censored using \eqref{censoringrule}.
  \State Estimator receives $(z_n,c_n,\mathbf{x}_n)$.
  \State Parameter $\boldsymbol{\theta}$ is updated via \eqref{SGD} and~\eqref{beta}.
  \EndFor
  \end{algorithmic}\label{SAMLEalg}
\end{algorithm}

Observe that when the $n$-th datum is not censored $(c_n=0)$, the second summand in the right-hand side (RHS) of \eqref{beta} vanishes, and \eqref{SGD} reduces to an ordinary LMS update. When $c_n=1$, the first summand disappears, and the update in \eqref{SGD} exploits the fact that the unavailable $y_n$ lies in a known interval $(|y_n-\mathbf{x}_n^T\hat{\boldsymbol{\theta}}_K|\leq{\tau_n\sigma})$, information that would have been ignored by an ordinary LMS algorithm.

Since the SA-MLE is in fact a Robbins-Monroe iteration on the sequence $\{\mathbf{g}(\boldsymbol{\theta})\}_{n=1}^D$, it inherits related convergence properties.  Specifically, by selecting $\mu_n=1/(nM)$ (for an appropriate $M$), the SA-MLE algorithm is asymptotically efficient and Gaussian~\cite[pg. 197]{young1974classification}. Performance guarantees also hold with finite samples. Indeed, with $D$ finite, the \emph{regret} attained by iterates $\{\boldsymbol{\theta}_n\}$ against a vector $\boldsymbol{\theta}$ is defined as
\begin{equation}\label{eq:regret}
R(D):=\sum_{n=1}^{D}\left[\ell_n(\boldsymbol{\theta}_n)-\ell_n(\boldsymbol{\theta})\right].
\end{equation}
Selecting $\mu$ properly, Algorithm~\ref{SAMLEalg} can afford bounded regret as asserted next; see Appendix for the proof.

\begin{proposition}\label{pro:regret}
Suppose $\|\mathbf{x}_n\|_2\le\bar{x}$ and $|\beta_n(\boldsymbol{\theta})|\le\bar\beta$ for $n=1,\ldots,D$, and let $\boldsymbol{\theta}^{\ast}$ be the minimizer of \eqref{MLE}. By choosing $\mu=\|\boldsymbol{\theta}^\ast-\hat{\boldsymbol{\theta}}_K\|_2/(\sqrt{2D}\bar\beta\bar{x})$, the regret of the SA-MLE satisfies
\begin{equation*}\label{eq4prop1}
R(D)\le \sqrt{2D}\|\boldsymbol{\theta}^\ast-\hat{\boldsymbol{\theta}}_K\|_2\bar{x}\bar{\beta}~.
\end{equation*}
\end{proposition}

Proposition~\ref{pro:regret} assumes bounded $\mathbf{x}_n$'s and noise. Although the latter is not satisfied by e.g., the Gaussian distribution, appropriate bounds ensure that \eqref{eq4prop1} holds with high probability.

\subsection{Second-Order SA-MLE}\label{subsec:2DSAMLE}
If extra complexity can be afforded, one may consider incorporating second-order information in the SA-MLE update to improve its performance. In practice, 
this is possible by replacing scalar with matrix step-sizes $\mathbf{M}_n$. Thus, the first-order stochastic gradient descent (SGD) update in~\eqref{SGD} is modified as follows 
\begin{equation}\label{SGDmat}
\boldsymbol{\theta}_{n}:=\boldsymbol{\theta}_{n-1}-\mathbf{M}_n^{-1}\mathbf{g}_n(\boldsymbol{\theta}_{n-1}).
\end{equation}
When solving $\min_{\boldsymbol{\theta}}~ \mathbb{E}[\ell_n(\boldsymbol{\theta})]$ using a second-order SA iteration, a desirable Newton-like matrix step size is $\mathbf{M}_n=\mathbb{E}[\nabla^2\ell_n(\boldsymbol{\theta}_{n})]$. Given that the latter requires knowing the average Hessian that is not available in practice, it is commonly surrogated by its sample-average $(1/n)\sum_{i=1}^{n} \nabla^2\ell_i(\boldsymbol{\theta}_{i})$~\cite{bertsekas2015convex}. To this end, note first that  $\nabla^2\ell_n(\boldsymbol{\theta})=\gamma_n(\boldsymbol{\theta})\mathbf{x}_n\mathbf{x}_n^T$, where
\begin{align}\label{gamma}
\gamma_n(\boldsymbol{\theta})&:=-\frac{(1-c_n)}{\sigma^2}-\frac{c_n}{\sigma^2}\Bigg[\left(\frac{\phi\left(z_n^u(\boldsymbol{\theta})\right)-\phi\left(z_n^l(\boldsymbol{\theta})\right)}{Q\left(z_n^u(\boldsymbol{\theta})\right)-Q\left(z_n^l(\boldsymbol{\theta})\right)}\right)^2\nonumber\\
	&-\frac{z_n^u(\boldsymbol{\theta}\phi\left(z_n^u(\boldsymbol{\theta})\right)-z_n^l(\boldsymbol{\theta}\phi\left(z_n^l(\boldsymbol{\theta})\right)}{Q\left(z_n^u(\boldsymbol{\theta})\right)-Q\left(z_n^l(\boldsymbol{\theta})\right)}\Bigg].
\end{align}
Due to the rank-one update $\mathbf{M}_n=((n-1)/n)\mathbf{M}_{n-1}+(1/n)\gamma_{n-1}(\boldsymbol{\theta}_{n-1})$ $\mathbf{x}_{n-1}\mathbf{x}_{n-1}^T$, the matrix step size $\mathbf{C}_n:=\mathbf{M}_n^{-1}$ can be obtained efficiently using the matrix inversion lemma as 
\begin{equation}\label{Cupdate}
\mathbf{C}_n=\frac{n}{n-1}\left(\mathbf{C}_{n-1}-\frac{\mathbf{C}_{n-1}\mathbf{x}_n\mathbf{x}_n^T
\mathbf{C}_{n-1}}{(n-1)\gamma_n^{-1}(\boldsymbol{\theta}_{n-1})+\mathbf{x}_n^T\mathbf{C}_{n-1}\mathbf{x}_n} \right).
\end{equation}
Similar to its first-order counterpart, the algorithm is initialized by the preliminary estimate $\boldsymbol{\theta}_0=\hat{\boldsymbol{\theta}}_K$, and $\mathbf{C}_0 = \sigma^2(\mathbf{X}_K^T\mathbf{X}_K)^{-1}$. The second-order SA-MLE method is summarized as Algorithm~\ref{alg:2dSAMLE}, while the numerical tests of Section~\ref{SAMLE} confirm its faster convergence at the cost of $\mathcal{O}(p^2)$ complexity per update.
\begin{algorithm}[t]
	\caption{Second-order SA-MLE}\label{alg:2dSAMLE}
	\begin{algorithmic}
		  \State Initialize $\boldsymbol{\theta}_0$ as the LSE $\hat{\boldsymbol{\theta}}_{K}$ in \eqref{LSE}.
		\State Initialize $\mathbf{C}_0=\sigma^2(\mathbf{X}_K^T\mathbf{X}_K)^{-1}$.
		\For {$n=1:D$}
		\State Measurement $y_n$ is possibly censored using \eqref{censoringrule}.
		\State Estimator receives $(z_n,\mathbf{x}_n,c_n)$.
		\State Compute $\gamma_n(\boldsymbol{\theta}_{n-1})$ from~\eqref{gamma}.
		\State Update matrix step size from \eqref{Cupdate}.
		\State Update parameter estimate as in \eqref{SGDmat}.
		\EndFor
	\end{algorithmic}
\end{algorithm}
\subsection{Controlling Data Reduction via NAC}\label{subsec:NACprob}
To apply the NAC rule of \eqref{censoringrule} for data reduction at a controllable rate, a relation between thresholds $\{\tau_n\}$ and the censoring rate must be derived. Furthermore, prior knowledge of the problem at hand (e.g., observations likely to contain outliers) may dictate a specific pattern of censoring probabilities $\{\pi_n^{\ast}\}_{n=1}^D$. If $d$ is the number of uncensored data after NAC is applied on a dataset of size $D\geq{d}$, then $(D-d)/D$ is the censoring ratio. Since $\{y_n\}$ are generated randomly according to \eqref{model}, it is clear that $d$ is itself a random variable. The analysis is thus focused on the average censoring ratio
\begin{equation}\label{ratio}
\bar{c}:=\mathbb{E}\left[\frac{D-d}{D}\right]=\frac{1}{D}\sum_{n=1}^{D}\mathbb{E}[c_n]=\frac{1}{D}\sum_{n=1}^{D}\pi_n
\end{equation}
where $\pi_n:=\Pr(c_n=1)$ is the probability of censoring datum $n$, that as a function of $\tau_n$ is given by [cf.~\eqref{censoringrule}]
\begin{align}\label{eq:pi}
\pi_n(\tau_n)&= \Pr\{-\tau_n\sigma\leq{y_n-\hat{y}_n}\leq{\tau_n\sigma}\}\nonumber\\
&=\Pr\{-\tau_n \leq \frac{\mathbf{x}_n^T(\boldsymbol{\theta}_o-\hat{\boldsymbol{\theta}}_{K})+v_n}{\sigma}\leq \tau_n \}.
\end{align}
By the properties of the LSE, $\hat{\boldsymbol{\theta}}_{K} \sim{\mathcal{N}(\boldsymbol{\theta}_o,\sigma^2(\mathbf{X}_K^T\mathbf{X}_K)^{-1})}$, it follows that 
\begin{equation*}
\frac{\mathbf{x}_n^T(\boldsymbol{\theta}_o-\hat{\boldsymbol{\theta}}_{K})+v_n}{\sigma}\sim \mathcal{N}\left(0,\mathbf{x}_n^T(\mathbf{X}_K^T\mathbf{X}_K)^{-1}\mathbf{x}_n+1\right).
\end{equation*} 
Thus, the censoring probabilities in \eqref{eq:pi} simplify to
\begin{equation}\label{tauexact}
\pi_n(\tau_n)=1-2Q\left({\tau_n\left[\mathbf{x}_n^T(\mathbf{X}_K^T\mathbf{X}_K)^{-1}\mathbf{x}_n+1\right]^{-1/2}}\right).
\end{equation}
Solving \eqref{tauexact} for $\tau_n$, one arrives for a given ${\pi}^{\star}_n=\pi_n(\tau_n^{\star})$ at
\begin{equation}\label{taulookup}
\tau_n^{\star}=\left[\mathbf{x}_n^T(\mathbf{X}_K^T\mathbf{X}_K)^{-1}\mathbf{x}_n+1\right]^{1/2}Q^{-1}\left(\frac{1-{\pi}^{\star}_n}{2}\right)\:.
\end{equation}
Hence, for a prescribed $\bar{c}$, one can select a desired censoring probability pattern $\{\pi^{\star}_n\}_{n=1}^{D}$ to satisfy~\eqref{ratio}, and corresponding $\{\tau_n^{\star}\}_{n=1}^D$ in accordance with~\eqref{taulookup}.

\begin{figure}[h!]
	\centering
	\subfigure[]{
		\includegraphics[width=1\linewidth, height=2 in]{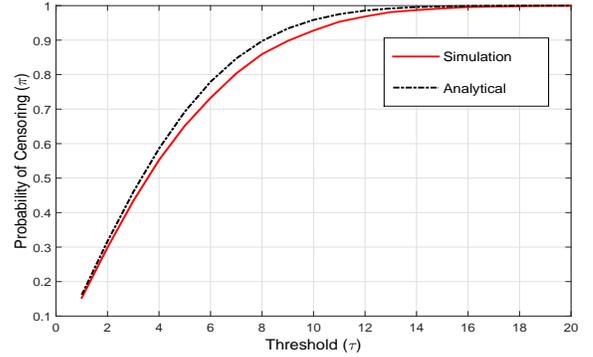}	
				\label{fig:prob1}
	}
	\subfigure[]{
		\includegraphics[width=1\linewidth, height=2 in]{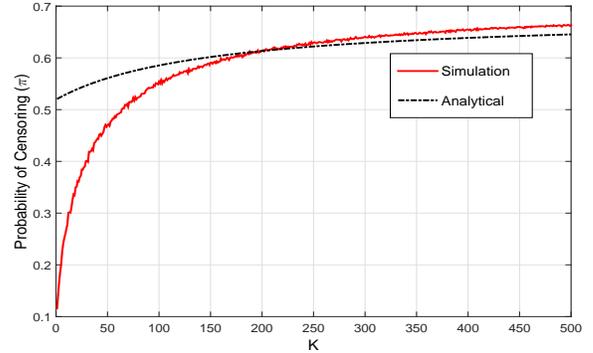}
				\label{fig:prob2}
	}
	\caption{a) Censoring probability for varying threshold $(p=100,K=200)$. b) Censoring probability for varying $K$ $(p=100,\tau=1)$.}
\end{figure}

The threshold selection \eqref{taulookup} requires knowledge of all $\{\mathbf{x}_n\}_{n=1}^D$. In addition, implementing \eqref{taulookup} for all $D$ observations, requires $\mathcal{O}(Dp^2)$ computations that may not be affordable for $D\gg p$. To deal with this, the ensuing simple  threshold selection rule is advocated. Supposing that $\{\mathbf{x}_n\}_{n=1}^D$ are generated i.i.d. according to some unknown distribution with known first- and second-order moments, a relation between a target \emph{common} censoring probability ${\pi}^{\star}$ and a common threshold $\tau$ can be obtained in closed form. Assume without loss of generality that $\mathbb{E}\left[\mathbf{x}_n\right]=\mathbf{0}$, and let $\mathbb{E}\left[\mathbf{x}_n\mathbf{x}_n^T\right]=\mathbf{R}_{x}$ and  $\boldsymbol{\zeta}_K:=(\boldsymbol{\theta}_o-\hat{\boldsymbol{\theta}}_K)/\sigma\sim \mathcal{N}(\mathbf{0},(\mathbf{X}_K^T\mathbf{X}_K)^{-1})$.
For sufficiently large $K$, it holds that $(\mathbf{X}_K^T\mathbf{X}_K)^{-1} \approx\mathbf{R}_x^{-1}/K$, and thus $\boldsymbol{\zeta}_K\sim \mathcal{N}(\mathbf{0},\mathbf{R}_x^{-1}/K)$. Next, using the standardized Gaussian random vector $\mathbf{u}\sim\mathcal{N}(\mathbf{0},\mathbf{I}_p)$, one can write $\boldsymbol{\zeta}_K=\mathbf{R}_x^{-1/2}\mathbf{u}/\sqrt{K}$. Also, with an independent zero-mean random vector $\mathbf{u}_n$ with $\mathbb{E}[\mathbf{u}_n\mathbf{u}_n^T]=\mathbf{I}_p$, it is also possible to 
express $\mathbf{x}_n=\mathbf{R}_x^{1/2}\mathbf{u}_n$, which implies 
$\mathbf{x}_n^T\boldsymbol{\zeta}_K=\mathbf{u}_n^T\mathbf{u}/\sqrt{K}$. 
By the central limit theorem (CLT), $\mathbf{u}_n^T\mathbf{u}$ converges 
in distribution to $\mathcal{N}(0,p)$ as the inner dimension of the two vectors $p$ grows; 
thus, $\mathbf{x}_n^T\boldsymbol{\zeta}_K\sim\mathcal{N}(0,p/K)$. 
Under this approximation, it holds that
\begin{align}\label{NACprob}
\pi_n\approx{\pi}&= Q\left({-\frac{\tau}{\sqrt{p/K+1}}}\right) -Q\left({\frac{\tau}{\sqrt{p/K+1}}}\right)\nonumber\\
&= 1-2Q\left(\frac{\tau}{\sqrt{p/K+1}}\right),~~n=1, \ldots, D.
\end{align}
As expected, due to the normalization by $\sigma$ in \eqref{censoringrule}, 
${\pi}$ does not depend on $\sigma$. Interestingly, it does not depend on 
$\mathbf{R}_{x}$ either. Having expressed $\pi$ as a function of $\tau$, the latter can be tuned to achieve the desirable data reduction. Following the law of large numbers and given parameters $p$ and $K$, to achieve an average censoring ratio of $\bar{c}={\pi}^{\star}=(D-d)/D$, the threshold can be set to
\begin{equation}\label{taurule}
\tau=\sqrt{1+p/K}\: Q^{-1}\left( \tfrac{1-{\pi}^{\star}}{2} \right).
\end{equation}
Figure~\ref{fig:prob1} depicts $\pi$ as a function of $\tau$ for $p=100$ and $K=200$. 
Function~\eqref{NACprob} is compared with the simulation-based estimate of $\pi_n$ 
using 100 Monte Carlo runs, confirming that \eqref{NACprob} offers a reliable 
approximation of ${\pi}$, which improves as $p$ grows. However, for the approximation 
$(\mathbf{X}_K^T \mathbf{X}_K)^{-1}\approx\mathbf{R}_x^{-1}/{K}$ to be accurate, $K$ 
should be large too. Figure~\ref{fig:prob2} shows the probability of censoring 
for varying $K$ with fixed $p=100$ and $\tau=1$. Approximation~\eqref{NACprob} 
yields a reliable value for $\pi$ for as few as $K\approx200$ preliminary data.

\section{Big Data Streaming Regression with AC}\label{sec:ACstreaming}
The NAC-based algorithms of Section~\ref{sec:NAConline} emerge in a wide range of applications for which censoring occurs naturally as part of the data acquisition process; see e.g., the Tobit model in economics~\cite{tobit}, and survival data analytics in~\cite{bio2008evers}. Apart from these applications where data are inherently censored, our idea is to \emph{employ censoring deliberately} for data reduction. Leveraging NAC for data reduction decouples censoring from estimation, and thus eliminates the need for obtaining further information. However, one intuitively expects improved performance with a \emph{joint censoring-estimation} design.  

In this context, first- and second-order sequential algorithms will be developed in this section for the AC in~\eqref{ACcensoringrule}. Instead of $\hat{\boldsymbol{\theta}}_K$, AC is performed using the latest estimate of $\boldsymbol{\theta}$. Apart from being effective in handling streaming data, AC can markedly lower the complexity of a \emph{batch} LS problem. Section~\ref{subsec:ACLMS} introduces an AC-based LMS algorithm for large-scale streaming regressions, while Section~\ref{subsec:ACRLS} puts forth an AC-based recursive least-squares (RLS) algorithm as a viable alternative to random projections and sampling.

\subsection{AC-LMS}\label{subsec:ACLMS}
A first-order AC-based algorithm is presented here, inspired by the celebrated LMS algorithm. Originally developed for adaptive filtering, LMS is well motivated for low-complexity online estimation of (possibly slow-varying) parameters. Given $(y_n,\mathbf{x}_n)$, LMS entails the simple update 
\begin{equation}\label{eq:LMS}
\boldsymbol{\theta}_{n}=\boldsymbol{\theta}_{n-1}+\mu\mathbf{x}_n e_n(\boldsymbol{\theta}_{n-1})
\end{equation}
where $e_n(\boldsymbol{\theta}):=y_n-\mathbf{x}_n^T\boldsymbol{\theta}$ can be viewed as the innovation of $y_n$, since $\hat{y}_n=\mathbf{x}_n^T\boldsymbol{\theta}_{n-1}$ is the prediction of $y_n$ given $\boldsymbol{\theta}_{n-1}$. LMS can be regarded as an SGD method for 
$\min_{\boldsymbol{\theta}} ~ \mathbb{E}[f_n(\boldsymbol{\theta})]$, 
where the instantaneous cost is $f_n(\boldsymbol{\theta})=e_n^2(\boldsymbol{\theta})/2$. 

To derive a first-order method for online censored regression, consider minimizing $\mathbb{E}[f_n^{(\tau)}(\boldsymbol{\theta})]$ with the instantaneous cost selected as the \emph{truncated} quadratic function
\begin{align}\label{inst}
f_n^{(\tau)}(\boldsymbol{\theta}):=
\left\{\begin{array}{ll}
\frac{e_n^2(\boldsymbol{\theta})-\tau_n^2\sigma^2}{2}	&,~|e_n(\boldsymbol{\theta})|\geq \tau_n\sigma\\
0	&,~|e_n(\boldsymbol{\theta})|< \tau_n\sigma
\end{array}\right.
\end{align}
for a given $\tau_n>0$. For the sake of analysis, a common threshold will be adopted; that is, $\tau_n=\tau$ $\forall n$. The truncated cost can be also expressed as $f_n^{(\tau)}(\boldsymbol{\theta})=\max\{0,(e_n^2(\boldsymbol{\theta})-\tau^2\sigma^2)/2\}$. Being the pointwise maximum of two convex functions, $f_n^{(\tau)}(\boldsymbol{\theta})$ is convex, yet not everywhere differentiable. From standard rules of subdifferential calculus, its subgradient is
\begin{equation}
\partial {f_n^{(\tau)}(\boldsymbol{\theta})}=\left\{\begin{array}{ll}
-\mathbf{x}_n e_n(\boldsymbol{\theta})&,~|e_n(\boldsymbol{\theta})|>\tau\sigma\\
\mathbf{0}&,~|e_n(\boldsymbol{\theta})|<\tau\sigma\\
\{-\varphi\mathbf{x}_n e_n(\boldsymbol{\theta}):0\leq \varphi \leq 1\}&,~|e_n(\boldsymbol{\theta})|=\tau\sigma
\end{array}\right . \:. \nonumber
\end{equation}
An SGD iteration for the instantaneous cost in \eqref{inst} with $\tau_n=\tau$, 
performs the following AC-LMS update per datum $n$
\begin{equation}\label{eq:ACLMS}
\boldsymbol{\theta}_{n} := \left\{\begin{array}{ll}
\boldsymbol{\theta}_{n-1}+\mu\mathbf{  x}_n e_n(\boldsymbol{\theta}_{n-1}) &,~|e_n(\boldsymbol{\theta}_{n-1})|\geq \tau\sigma\\
\boldsymbol{\theta}_{n-1} &,~\textrm{otherwise}
\end{array}\right.
\end{equation}
where $\mu>0$ can be either constant for tracking a time-varying parameter, or, 
diminishing over time for estimating a time-invariant $\boldsymbol{\theta}_o$. 
Different from SA-MLE, the AC-LMS does not update $\boldsymbol{\theta}$ if datum $n$ is censored. The intuition is that if $y_n$ can be closely predicted by $\hat{y}_n :=\mathbf{x}_n^T \boldsymbol{\theta}_{n-1}$, then $(y_n,\mathbf{x}_n)$ can be censored (small innovation is indeed `not much informative'). Extracting interval information through a likelihood function as in Algorithm~\ref{SAMLEalg} appears to be challenging here. This is because unlike NAC, the AC data $\{z_n\}_{n=1}^{D}$ are dependent across time. 

Interestingly, upon invoking the ``independent-data assumption'' of SA~\cite{young1974classification}, following the same steps as in 
Section~\ref{sec:NAConline}, and substituting $\hat{\boldsymbol{\theta}}_K=\boldsymbol{\theta}_{n-1}$ into \eqref{beta}, the interval information term is eliminated. This is a strong indication that interval information from censored observations may be completely ignored without the risk of introducing bias. Indeed, one of the implications of the ensuing Proposition~\ref{pro:converge_aclms} is that the AC-LMS is asymptotically unbiased. Essentially, in AC-LMS as well as in the AC-RLS to be introduced later, both $\mathbf{x}_n$ and $y_n$ are censored -- an important feature effecting further data reduction and lowering computational complexity of the proposed AC algorithms. The mean-square error (MSE) performance of AC-LMS is established in the next proposition proved in the Appendix.

\begin{proposition}\label{pro:converge_aclms}
Assume $\mathbf{x}_n$'s are generated i.i.d. with $\mathbb{E}\left[\mathbf{x}_n\right]=\mathbf{0}$, $\mathbb{E}\left[\mathbf{x}_n\mathbf{x}_n^T\right]=\mathbf{R}_{x}$, $\mathbb{E}\left[\mathbf{x}_n^T\mathbf{x}_n\mathbf{x}_n^T\right]=\mathbf{0}^T$,  and $\mathbb{E}\left[\left(\mathbf{x}_n\mathbf{x}_n^T\right)^2\right]=\mathbf{R}_{x}^2$, while observations $y_n$ are obtained according to model \eqref{model}. For a diminishing  $\mu_n=\mu/n$ with $\mu=2/\alpha$, initial estimate $\boldsymbol{\theta}_1$, and censoring-controlling threshold $\tau$, the AC-LMS in~\eqref{eq:ACLMS} yields an estimate $\boldsymbol{\theta}_n$ with MSE bounded as
\begin{equation*}
\mathbb{E}\left[{\|{\boldsymbol{\theta}_n-\boldsymbol{\theta}_o}\|_2^2}\right] \leq{\frac{e^{4L^2/\alpha^2}}{n^2}}\left(\|{\boldsymbol{\theta}_1-\boldsymbol{\theta}_o}\|_2^2+\frac{\Delta}{L^2}\right)+\frac{8\Delta\log{n}}{\alpha^2n}
\end{equation*} 
where $\alpha :=2Q(\tau)\lambda_{\min}(\mathbf{R}_x)$, $\Delta := 2\mathrm{tr}(\mathbf{R}_x)\sigma^2(1-Q(\tau)$ $+\tau p(\tau))$, and $L^2 :=\lambda_{\max}\left(\mathbf{R}_x^2\right)$. Further, for $\mu$ $<\alpha/(16L^2)$, AC-LMS converges exponentially to a bounded error
\begin{align*}
\mathbb{E}\left[{\|{\boldsymbol{\theta}_n-\boldsymbol{\theta}_o}\|_2^2}\right] &\leq 2\exp\left(-\left(\frac{\alpha\mu}{4}-4L^2\mu^2\right)n-4L^2\mu^2\right)\\
	&\times\left(\|{\boldsymbol{\theta}_1-\boldsymbol{\theta}_o}\|_2^2+\frac{\Delta}{L^2}\right)+\frac{4\mu\Delta}{\alpha}.
\end{align*} 
\end{proposition}

Proposition~\ref{pro:converge_aclms} asserts that AC-LMS achieves a bounded MSE. It also links MSE with the AC threshold $\tau$ that can be used to adjust the censoring probability. Closer inspection reveals that the MSE bound decreases with $\tau$. In par with intuition, lowering $\tau$ allows the estimator to access more data, thus enhancing estimation performance at the price of increasing the data volume processed.

\subsection{AC-RLS}\label{subsec:ACRLS}
A second-order AC algorithm is introduced here for the purpose of sequential estimation and dimensionality reduction. It is closely related to the RLS algorithm, which per time $n$ implements the updates; see e.g.,~\cite{slavakis2014stochastic} 
\begin{subequations}
\begin{align}\label{RLS}
\mathbf{C}_n &= \frac{n}{n-1}
\left[\mathbf{C}_{n-1} 
- \frac{\mathbf{C}_{n-1}\mathbf{x}_n\mathbf{x}_n^T\mathbf{C}_{n-1}}{n-1+\mathbf{x}_n^T\mathbf{C}_{n-1}\mathbf{x}_n}
\right]\\
\boldsymbol{\theta}_{n}  &= \boldsymbol{\theta}_{n-1} + \frac{1}{n}\mathbf{C}_n\mathbf{x}_n(y_n-\mathbf{x}_n^T\boldsymbol{\theta}_{n-1})
\end{align}
\end{subequations}
where $\mathbf{C}_n$ is the sample estimate for $\mathbf{R}_x^{-1}$ and is typically initialized to  $\mathbf{C}_0=\epsilon \mathbf{I}$, for some small positive $\epsilon$, e.g.,~\cite{kay93book}. The RLS estimate at time $n$ can be also obtained as  
\begin{align}\label{RLSexact}
\boldsymbol{\theta}_n=\arg\min_{\boldsymbol{\theta}}{\sum_{i=1}^{n}{(y_i-\mathbf{x}_i^T\boldsymbol{\theta})^2}+\epsilon\|\boldsymbol{\theta}\|_2^2}.
\end{align}
The bias introduced by the arbitrary choice of $\mathbf{C}_0$ vanishes asymptotically in $n$, while the RLS iterates converge to the batch LSE. RLS can be viewed as a second-order SGD method of the form $\boldsymbol{\theta}_n=\boldsymbol{\theta}_{n-1}-\mathbf{M}_n^{-1}\nabla{f_n(\boldsymbol{\theta}_{n-1})}$ for the quadratic cost $f_n(\boldsymbol{\theta})=e_n^2(\boldsymbol{\theta})/2$. In this instance of SGD, the ideal matrix step size $\mathbf{M}_n=\mathbb{E}[\nabla^2f_n(\boldsymbol{\theta}_{n-1})]=\mathbb{E}\left[(1-c_n)\mathbf{x}_n\mathbf{x}_n^T\right]$ is replaced by its running estimate $(1/n)\mathbf{C}_n^{-1}$; see e.g.,~\cite{bertsekas2015convex}.

To obtain a second-order counterpart of AC-LMS, we replace the quadratic instantaneous cost of RLS with the truncated quadratic in~\eqref{inst}. The matrix step-size is further surrogated by
\begin{equation*}\label{eq:mss}
\mathbf{M}_n= \frac{1}{n}\sum\limits_{i=1}^{n}(1-c_i)\mathbf{x}_i\mathbf{x}_i^T=\frac{n-1}{n}\mathbf{M}_{n-1}+\frac{1}{n}(1-c_n)\mathbf{x}_n\mathbf{x}_n^T.
\end{equation*}
Applying the matrix inversion lemma to find $\mathbf{M}_n^{-1}$ yields the next AC-RLS updates
\begin{subequations}\label{ACRLS}
\begin{align}
\mathbf{C}_n &= \frac{n}{n-1}
\left[\mathbf{C}_{n-1}-\frac{(1-c_n)\mathbf{C}_{n-1} \mathbf{x}_n\mathbf{x}_n^T\mathbf{C}_{n-1}}{n-1+\mathbf{x}_n^T\mathbf{C}_{n-1}\mathbf{x}_n} 
\right]\label{ACRLS:C}\\
\boldsymbol{\theta}_{n}  &= \boldsymbol{\theta}_{n-1} + \frac{1-c_n}{n}\mathbf{C}_n\mathbf{x}_n(y_n-\mathbf{x}_n^T\boldsymbol{\theta}_{n-1})\label{ACRLS:theta}
\end{align}
\end{subequations}
where $c_n$ is decided by \eqref{ACcensoringrule}. For $c_n=1$, the parameter vector is not updated, while costly updates of $\mathbf{C}_n$ are also avoided. In addition, different from the iterative expectation-maximization algorithm in \cite{tsp2013youxie}, AC-RLS skips completely covariance updates. Its performance is characterized by the following proposition shown in the Appendix.

\begin{proposition}\label{pro:converge_acrls}
If $\mathbf{x}_n$'s are i.i.d. with $\mathbb{E}\left[\mathbf{x}_n\right]=\mathbf{0}$ and $\mathbb{E}\left[\mathbf{x}_n\mathbf{x}_n^T\right]=\mathbf{R}_{x}$, while observations $y_n$ adhere to the model in \eqref{model}, then for $\boldsymbol{\theta}_1=\mathbf{0}$ and constant $\tau$, there exists $k>0$ such that AC-RLS estimates $\boldsymbol{\theta}_n$ yield bounded MSE 
\begin{equation*}
\frac{1}{n}\mathrm{tr}\left(\mathbf{R}_x^{-1}\right)\sigma^2\leq\mathbb{E}\left[{\|\boldsymbol{\theta}_n-\boldsymbol{\theta}_o\|_2^2}\right] \leq{\frac{1}{n}\frac{\mathrm{tr}\left(\mathbf{R}_x^{-1}\right)\sigma^2}{2Q(\tau)}},~~\forall n\geq k.
\end{equation*} 
\end{proposition}

As corroborated by Proposition~\ref{pro:converge_acrls}, the AC-RLS estimates are guaranteed to converge to $\boldsymbol{\theta}_o$ for any choice of $\tau$.  Overall, the novel AC-RLS algorithm offers a computationally-efficient and accurate means of solving large-scale LS problems encountered with Big Data applications.
 
 \begin{algorithm}[t]
  \caption{Adaptive-Censoring (AC)-RLS}\label{alg:ACRLS}
  \begin{algorithmic}
  \State Initialize $\boldsymbol{\theta}_0=\mathbf{0}$ and $\mathbf{C}_0=\epsilon\mathbf{I}$.
  \For {$n=1:D$ }
  \If{$\left|{y_n-\mathbf{x}_n^T\boldsymbol{\theta}_{n-1}}\right|\geq\tau\sigma$}
  \State Estimator receives $(y_n,\mathbf{x}_n)$ while $c_n=0$.
  \State Update inverse sample covariance from \eqref{ACRLS:C}.
  \State Update estimate from \eqref{ACRLS:theta}.
  \Else
  \State Estimator receives no information $(c_n=1)$.
  \State Propagate inverse covariance as $\mathbf{C}_n= \frac{n}{n-1}\mathbf{C}_{n-1}$.
  \State Preserve estimate $\boldsymbol{\theta}_{n}=\boldsymbol{\theta}_{n-1}$.
  \EndIf
  \EndFor
  \end{algorithmic}
\end{algorithm}
 
At this point, it is useful to contrast and compare AC-RLS with RP and random sampling methods that have been advocated as fast LS solvers~\cite{mahoney2011arxiv,mahoney2011ftml}. In practice, RP-based schemes first premultiply data $(\mathbf{y},\mathbf{X})$ with a random matrix $\mathbf{R}=\mathbf{HD}$, where $\mathbf{H}$ is a $D\times D$ Hadamard matrix and $\mathbf{D}$ is a diagonal matrix whose diagonal entries take values $\{-1/\sqrt{D},+1/\sqrt{D}\}$ equiprobably. Intuitively, $\mathbf{R}$ renders all rows of ``comparable importance'' (quantified by the leverage scores \cite{mahoney2011arxiv, mahoney2011ftml}), so that the ensuing random matrix $\mathbf{S}_d$ exhibits no preference in selecting uniformly a subset of $d$ rows. Then, the reduced-size LS problem can be solved as $\check{\boldsymbol{\theta}}_{d}=\arg\min_{\boldsymbol{\theta}}{\|{\mathbf{S}_d\mathbf{HD}(\mathbf{y}-\mathbf{X}\boldsymbol{\theta})}\|_2^2}$. For a general preconditioning matrix $\mathbf{HD}$, computing the products $\mathbf{HDy}$ and $\mathbf{HDX}$ requires a prohibitive number of $\mathcal{O}(D^2p)$ computations. This is mitigated by the fact that $\mathbf{H}$ has binary $\{+1,-1\}$ entries and thus multiplications can be implemented as simple sign flips. Overall, the RP method reduces the computational complexity of the LS problem from $\mathcal{O}(Dp^2)$ to $\scriptstyle{\mathcal{O}}$$(Dp^2)$ operations.

By setting $\tau=Q^{-1}(d/(2D))$, our AC-RLS Algorithm~\ref{alg:ACRLS} achieves an average  reduction ratio $d/D$ by scanning the observations, and selecting only the most informative ones. The same data ratio can be achieved more accurately by choosing a sequence of data-adaptive thresholds $\{\tau_n\}_{n=1}^D$, as described in the next subsection. As will be seen in Section~\ref{subsec:simACRLS}, AC-RLS achieves significantly lower estimation error compared to RP-based solvers. Intuitively, this is because unlike RPs that are based solely on $\mathbf{X}$ and are thus \emph{observation-agnostic}, AC extracts the most informative in terms of innovation subset of rows for a given problem instance $(\mathbf{y},\mathbf{X})$.

Regarding the complexity of AC-RLS, if the pair $(y_n,\mathbf{x}_n)$ is not censored, the cost of updating $\boldsymbol{\theta_n}$ and $\mathbf{C}_n$ is $\mathcal{O}(p^2)$ multiplications. For a censored datum, there is no such cost. Thus, for $d$ uncensored data the overall computational complexity is $\mathcal{O}(dp^2)$. Furthermore, evaluation of the absolute normalized innovation requires $\mathcal{O}(p)$ multiplications per iteration. Since this operation takes place at each of the $D$ iterations, there are $\mathcal{O}(Dp)$ computations to be accounted for. Overall, AC-RLS reduces the complexity of LS from $\mathcal{O}(Dp^2)$ to $\mathcal{O}(dp^2)+\mathcal{O}(Dp)$. Evidently, the complexity reduction is more prominent for larger model dimension $p$. For $p\gg{1}$, the second term may be neglected, yielding an $\mathcal{O}(dp^2)$ complexity for AC-RLS. 

A couple of remarks are now in order. 

\begin{rembold}
\rm{The novel AC-LMS and AC-RLS algorithms bear structural similarities to sequential set-membership (SM)-based estimation~\cite{bertsekas1971recursive,gollamudi1998set}. However, the model assumptions and objectives of the two are different. SM assumes 
that the noise distribution in~\eqref{model} has bounded support, which implies  that $\boldsymbol{\theta}_o$ belongs to a closed set. This set is sequentially identified by algorithms interpreted geometrically, while certain observations may be deemed redundant and thus discarded by the SM estimator. In our Big Data setup, an SA approach is developed to \emph{deliberately} skip updates of low importance for reducing complexity regardless of the noise pdf.}
\end{rembold}

\begin{rembold}
\rm{Estimating regression coefficients relying on ``most informative'' data is reminiscent 
of support vector regression (SVR), which typically adopts an $\epsilon$-insensitive cost (truncated $\ell_1$ error norm). SVR has well-documented merits in robustness as well as generalization capability, both of which are attractive for (even nonlinear kernel-based) prediction tasks~\cite{htf09book}. Solvers are typically based on nonlinear programming, and support vectors (SVs) are returned after \emph{batch} processing that does not scale well with the data size. Inheriting the merits of SVRs, the novel AC-LMS and AC-RLS can be viewed as returning ``causal SVs,'' which are different from the traditional (non-causal) batch SVs, but become available on-the-fly at complexity and storage requirements that are affordable for streaming Big Data. In fact, we conjecture that causal SVs returned by AC-RLS will approach their non-causal SVR counterparts if multiple passes over the data are allowed. Mimicking SVR costs, our AC-based schemes developed using the truncated $\ell_2$ cost~[cf. \eqref{inst}] can be readily generalized to their counterparts based on the truncated $\ell_1$ error norm. Cross-pollinating in the other direction, our AC-RLS iterations can be useful for online support vector machines capable of learning from streaming large-scale data with second-order closed-form iterations.}
\end{rembold}

\subsection{Controlling Data Reduction via AC}\label{subsec:ACcontrol}
A clear distinction between NAC and AC is that the latter depends on the estimation algorithm used. As a result, threshold design rules are estimation-driven rather than universal. In this section, threshold selection strategies are proposed for AC-RLS. Recall the average reduction ratio $\bar{c}$ in~\eqref{ratio}, and let 
$\boldsymbol{\zeta}_n:=(\boldsymbol{\theta}_o-\boldsymbol{\theta}_n)/\sigma\sim{\mathcal{N}(\mathbf{0},\mathbf{K}_n)}$ denote the normalized error at the $n-$th iteration. Similar to \eqref{ratio}--\eqref{eq:pi}, it holds that 
\begin{equation}\label{ACprob2}
\pi_n(\tau_n)= 1-2Q\left(\tau_n\left[\mathbf{x}_n^T\mathbf{K}_{n-1}\mathbf{x}_n+1\right]^{-  1/2}\right).
\end{equation}
For $n\gg{p}$, estimates $\boldsymbol{\theta}_n$ are sufficiently close to $\boldsymbol{\theta}_o$ and thus $\mathbf{K}_{n}\approx\mathbf{0}$. Then, the data-agnostic  $\tau_n\approx Q^{-1}(\frac{1-\pi_n}{2})$ attains an average censoring probability $\bar{\pi}$, while its asymptotic properties have been studied in~\cite{tsp2013youxie}. For finite data, this simple rule leads to under-censoring by ignoring appreciable values of $\mathbf{K}_n$, which can increase computational complexity considerably. This consideration motivates well the data-adaptive threshold selection rules designed next. 

AC-RLS updates can be seen as ordinary RLS updates on the subsequence of uncensored data. After ignoring the transient error due to initialization, it holds that $\mathbf{K}_n\approx\left[\sum_{i=1}^{n}(1-c_i)\mathbf{x}_i\mathbf{x}_i^T\right]^{-1}$. The term $\mathbf{x}_n^T\mathbf{K}_{n-1}\mathbf{x}_n$ is encountered as $\mathbf{x}_n^T \mathbf{C}_{n-1}\mathbf{x}_n/n$ in the updates of Alg.~\ref{alg:ACRLS}, but it is not computed for censored measurements. Nonetheless, $\mathbf{x}_n^T \mathbf{C}_{n-1}\mathbf{x}_n/n$ can be obtained at the cost of $p(p+1)$ multiplications per censored datum. Then, the exact censoring probability at AC-RLS iteration $n$ can be tuned to a prescribed $\pi^{\star}_n$ by selecting
\begin{equation}\label{ACtau}
\tau_n=\left(\mathbf{x}_n^T\mathbf{C}_{n-1}\mathbf{x}_n/n+1\right)^{1/2} Q^{-1}\left(\frac{1-\pi^{\star}_n}{2}\right).
\end{equation}
Given $\{\pi^{\star}_n\}_{n=1}^D$ satisfying \eqref{ratio}, an average censoring ratio of $(D-d)/D$ is thus achieved in a controlled fashion. 

Although lower than that of ordinary RLS, the complexity of AC-RLS using the threshold selection rule~\eqref{ACtau} is still $\mathcal{O}(Dp^2)$. To further lower complexity, a simpler rule is proposed that relies on averaging out the contribution of individual rows $\mathbf{x}_n^T$ in the censoring process. Suppose that $\mathbf{x}_n$'s are generated i.i.d. with $\mathbb{E}[\mathbf{x}_n]=\mathbf{0}$ and $\mathbb{E}[\mathbf{x}_n\mathbf{x}_n^T]=\mathbf{R}_x$. Similar to  Section~\ref{subsec:NACprob}, for $p$ sufficiently large the inner product $\mathbf{x}_n^T\boldsymbol{\zeta}_n$ is approximately Gaussian. It then follows that the a-priori error $e_n(\boldsymbol{\theta}_{n-1})=\sigma\mathbf{x}_n^T\boldsymbol{\zeta}_{n-1}+v_n$ is zero-mean Gaussian with variance $\sigma_{e_n}^2=\sigma^2\mathbb{E}\left[\mathbf{x}_n^T\boldsymbol{\zeta}_{n-1} \boldsymbol{\zeta}_{n-1}^T \mathbf{x}_n\right] + \sigma^2=\sigma^2\textrm{tr}\left(\mathbb{E}\left[\mathbf{x}_n\mathbf{x}_n^T\boldsymbol{\zeta}_{n-1} \boldsymbol{\zeta}_{n-1}^T\right]\right) + \sigma^2 =\sigma^2\mathrm{tr}\left(\mathbf{R}_{x}\mathbf{K}_{n-1}\right) + \sigma^2$, where the first equality follows from the independence of $\mathbf{x}_n^T\boldsymbol{\zeta}_{n-1}$ and $v_n$; and the third one from that of $\mathbf{x}_n$ with $\boldsymbol{\zeta}_{n-1}$. The censoring probability at time $n$ is then expressed as
\begin{equation*}
\pi_n=\Pr\{|e_n(\boldsymbol{\theta}_{n-1})|\leq{\tau\sigma}\}=1-2Q\left(\tau_n\frac{\sigma}{\sigma_{e_n}}\right).
\end{equation*}
To attain $\pi^{\star}_n$, the threshold per datum $n$ is selected as 
\begin{equation}\label{ACtau2}
\tau_n=\frac{\sigma_{e_n}}{\sigma}Q^{-1}\left(\frac{1-\pi^{\star}_n}{2}\right).
\end{equation}
It is well known that for large $n$, the RLS error covariance matrix $\mathbf{K}_n$ converges to $\frac{\sigma^2}{n}\mathbf{R}_{x}^{-1}$. Specifying $\{\pi^{\star}_n\}_{n=1}^D$ is equivalent to selecting an average number of $\sum_{i=1}^{n}(1-\pi^{\star}_i)$ RLS iterations until time $n$. Thus, the AC-RLS with controlled selection probabilities yields an error covariance matrix $\mathbf{K}_n\approx\left(\sum_{i=1}^{n}(1-\pi^{\star}_i)\right)^{-1}\sigma^2\mathbf{R}_{\mathbf{x}}^{-1}$. Combined with \eqref{ACtau2}, the latter leads to
\begin{equation*}
\sigma_{e_n}^2= \sigma^2p\left(\sum_{i=1}^{n-1}(1-\pi^{\star}_i)\right)^{-1} + \sigma^2.
\end{equation*}
Plugging $\sigma_{e_n}$ into \eqref{ACtau2} yields the simple threshold selection
\begin{equation}\label{ACtau3}
\tau_n=\left[p\left(\sum_{i=1}^{n-1}(1-\pi^{\star}_i)\right)^{-1} +1 \right]^{1/2}Q^{-1}\left(\frac{1-\pi^{\star}_n}{2}\right).
\end{equation}   
Unlike \eqref{ACtau}, where thresholds are decided online at an additional computational cost, \eqref{ACtau3} offers an off-line threshold design strategy for AC-RLS. Based on \eqref{ACtau3}, to achieve $\bar{c}={\pi}^{\star}=(D-d)/D$, thresholds are chosen as
\begin{equation}\label{ACtau4}
\tau_n=\left(\frac{p}{(n-1)(1-{\pi^{\star}})} +1\right)^{1/2}Q^{-1}\left(\frac{1-{\pi^{\star}}}{2}\right)
\end{equation}
which attains a constant $\pi^{\ast}$ across iterations. 

\subsection{Robust AC-LMS and AC-RLS}\label{subsec:robust}
AC-LMS and AC-RLS were designed to adaptively select data with relatively large innovation. This is reasonable provided that~\eqref{model} contains no outliers whose extreme values may give rise to large innovations too, and thus be mistaken for informative data. Our idea to gain robustness against outliers is to adopt the modified AC rule 
\begin{equation}\label{robustrule}
(c_n,c_n^o)=\left\{\begin{array}{ll}
(1,0)&,~|e_n(\boldsymbol{\theta}_{n-1})|<\sigma\tau\\
(0,0)&,~\tau\sigma\leq{|e_n(\boldsymbol{\theta}_{n-1})|}<{\tau_o\sigma}\\
(0,1)&,~|e_n(\boldsymbol{\theta}_{n-1})|\geq{\tau_o\sigma}
\end{array}\right..
\end{equation}   
Similar to \eqref{ACcensoringrule}, a nominal censoring variable $c_n$ is activated here too for observations with absolute normalized innovation less than $\tau$. To reveal possible outliers, a second censoring variable $c_n^o$ is triggered when the absolute normalized innovation exceeds threshold $\tau_o>\tau.$ 

Having separated data-censoring from outlier identification in \eqref{robustrule}, it becomes possible to robustify AC-LMS and AC-RLS against outliers. Towards this end, one approach is to completely ignore $y_n$ when $c_n^o=1$. Alternatively, the instantaneous cost function in \eqref{inst} can be modified to a truncated Huber loss (cf. \cite{huber1964robust})
\begin{equation}\label{instRobust}
f^{o}(e_n)=\left\{\begin{array}{ll}
0&,(c_n,c_n^o)=(1,0)\\
\left(\frac{1}{2}e_n^2-\frac{1}{2}\tau^2\sigma^2\right)&,(c_n,c_n^o)=(0,0)\\
\tau_o\sigma\left(|e_n|-\frac{3}{2}\tau_o^2\sigma^2-\frac{1}{2}\tau^2\sigma^2\right)&,(c_n,c_n^o)=(0,1)
\end{array}\right.\nonumber .
\end{equation}   
Applying the first-order SGD iteration on the cost $f^{o}(e_n)$, yields the robust (r) AC-LMS iteration
\begin{equation}\label{rACLMs}
\boldsymbol{\theta}_n=\boldsymbol{\theta}_{n-1}+\mu_n\mathbf{g}_n(\boldsymbol{\theta}_{n-1})
\end{equation} 
where
\begin{equation}
\mathbf{g}_n(\boldsymbol{\theta})=\left\{\begin{array}{ll}
\mathbf{0}&,~(c_n,c_n^o)=(1,0)\\
\mathbf{x}_n\left(y_n-\mathbf{x}_n^T\boldsymbol{\theta}_{n-1}\right)&,~(c_n,c_n^o)=(0,0)\\
\tau_o\sigma\mathbf{x}_n\sign\left(y_n-\mathbf{x}_n^T\boldsymbol{\theta}_{n-1}\right)&,~(c_n,c_n^o)=(0,1)
\end{array}\right..\nonumber
\end{equation}   
Similarly, the second-order SGD yields the rAC-RLS
\begin{subequations}\label{rACRLS}
\begin{align}
\boldsymbol{\theta}_{n}  &= \boldsymbol{\theta}_{n-1} + \frac{1}{n}{\mathbf{C}_n}\mathbf{g}_n(\boldsymbol{\theta}_{n-1})\\
\mathbf{C}_n &= \frac{n}{n-1}\left[
\mathbf{C}_{n-1} - \frac{(1-c_n)(1-c_n^o)\mathbf{C}_{n-1}\mathbf{x}_n\mathbf{x}_n^T\mathbf{C}_{n-1}}{n-1+\mathbf{x}_n^T\mathbf{C}_{n-1}\mathbf{x}_n}
\right].\label{rACRLS:C}
\end{align}
\end{subequations}
Observe that when $c_n^o=1$, only $\boldsymbol{\theta}_{n}$ is updated, and the computationally costly update of \eqref{rACRLS:C} is avoided.

\section{Numerical Tests}\label{sec:Sim}
\subsection{SA-MLE}\label{SAMLE}
\begin{figure}[t]
	\centering
	\centerline{\includegraphics[width=1\linewidth, height=2.1 in]{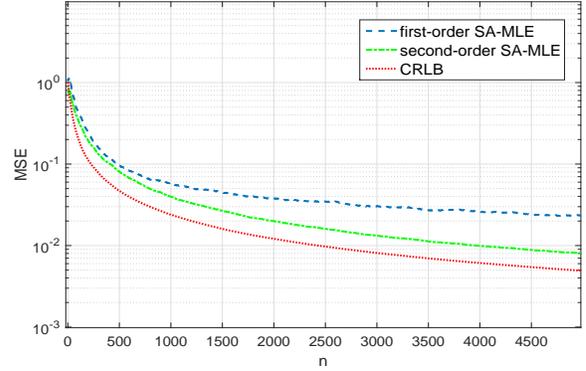}}
	\caption{Convergence of first- and second-order SA-MLE $(d/D=0.25)$ .}
	\vspace{0.1cm}
	\label{fig:SA_MLE_CRLB}
\end{figure}

The online SA-MLE algorithms presented in Section~\ref{sec:NAConline} 
are simulated using Gaussian data generated according to~\eqref{model}
with a time-invariant $\boldsymbol{\theta}_o\in\mathbb{R}^{p}$, where $p=30$, $\upsilon_n\sim\mathcal{N}(0,1)$ and
$\mathbf{x}_n\sim\mathcal{N}\left(\mathbf{0}_p,\mathbf{I}_p\right)$. The first $K=50$ observations are used to compute $\hat{\boldsymbol{\theta}}_{K}$. The first-and second-order SA-MLE algorithms
are then run for $D=5,000$ time steps. The NAC rule in~\eqref{censoringrule} was used with $\tau=1.5$ to censor approximately $75\%$ of the observations. Plotted in Fig.~\ref{fig:SA_MLE_CRLB} is the MSE $\mathbb{E}\Big[\|\boldsymbol{\theta}_o-\hat{\boldsymbol{\theta}}_n\|_2^2\Big]$ across time $n$, approximated by averaging over 100 Monte Carlo experiments. Also plotted is the Cramer-Rao lower bound (CRLB) of the observations, given by modifying the results of~\cite{tsp2012ggeric} to accommodate the NAC rule in~\eqref{censoringrule}. It can be inferred from the plot that the second-order SA-MLE exhibits markedly improved convergence rate compared to its first-order counterpart, at the price of minor increase in complexity. Furthermore, by performing a single pass over the data, the second-order SA-MLE performs close to the CRLB, thus offering an attractive alternative to the more computationally demanding batch Newton-based iterations in~\cite{tsp2013youxie} and~\cite{tsp2012ggeric}.

\begin{figure}[t]
\subfigure[]{
\centering
\centerline{\includegraphics[width=1\linewidth]{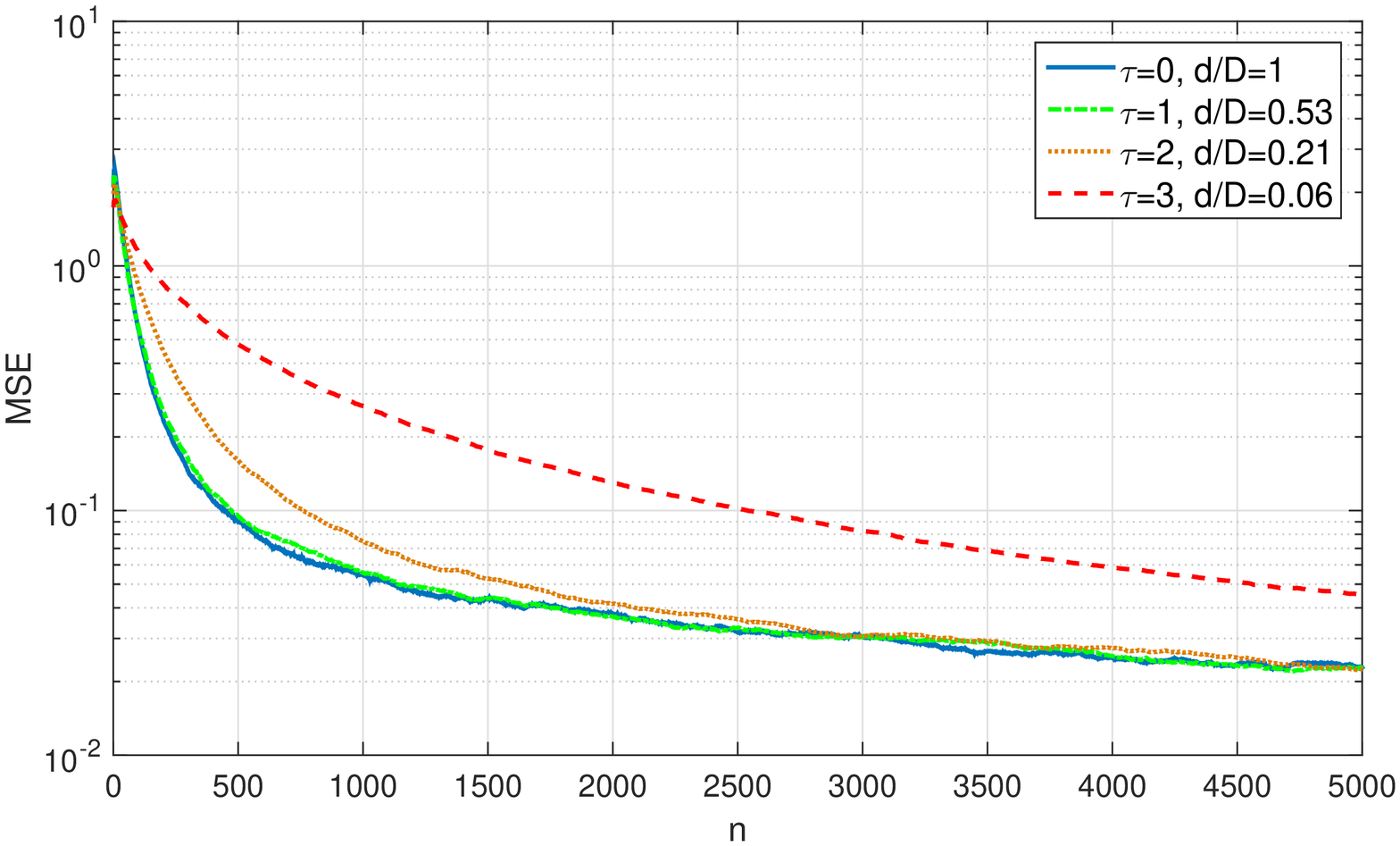}}			
\label{fig:SA_MLE}}
\subfigure[]{
\centering
\centerline{\includegraphics[width=1\linewidth]{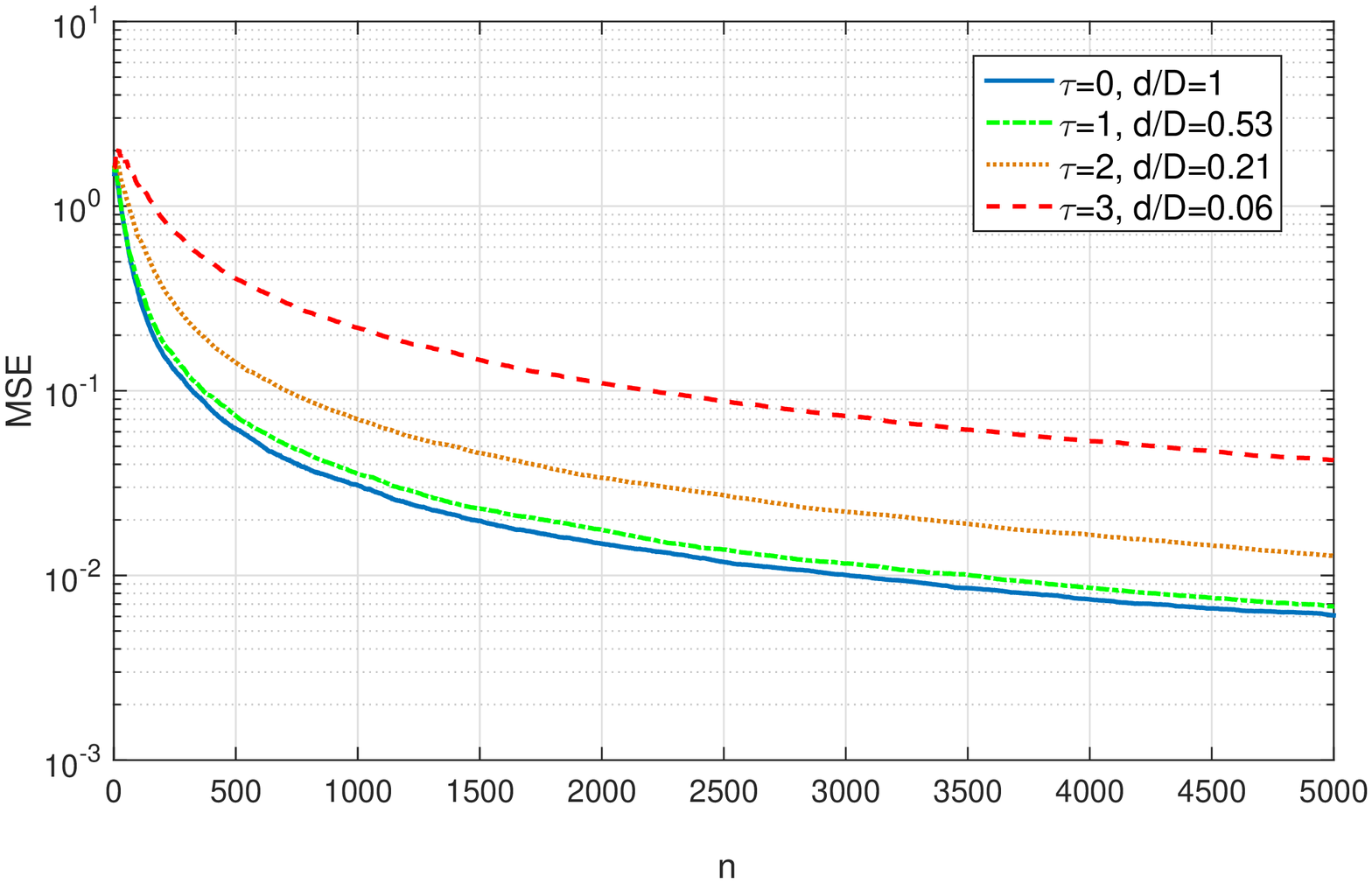}}
\label{fig:Sa_MLE2}}
\caption{Convergence of (a) first-order SA-MLE; and (b) second-order SA-MLE for different values of $\tau$.}
\end{figure}

To further evaluate the efficacy of the proposed methods, additional simulations were run for different levels of censoring by adjusting $\tau$. Plotted in Figs.~\ref{fig:SA_MLE} and~\ref{fig:Sa_MLE2} are the MSE curves of the first- and second-order SA-MLE respectively, for different values of $\tau$. Notice that censoring up to $50\%$ of the data (green solid curve) incurs negligible estimation error compared to the full-data case (blue solid curve). In fact, even when operating on data reduced by $95\%$ (red dashed curve) the proposed algorithms yield reliable online estimates. 

\subsection{AC-LMS comparison with Randomized Kaczmarz}

\begin{figure}[t]
\centering
\centerline{\includegraphics[width=0.85\linewidth, height=2 in]{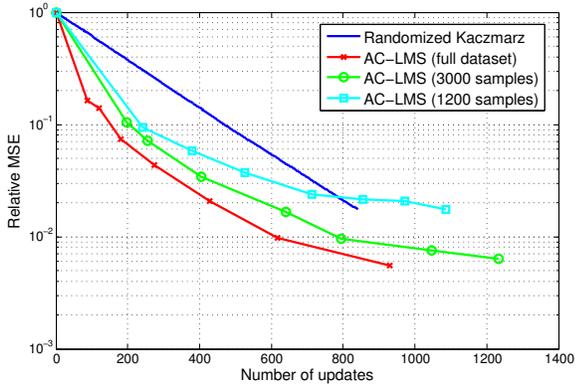}}
\caption{Relative MSE for AC-LMS and randomized Kaczmarz's algorithms.}
\label{fig:AC_LMS}
\end{figure}

The AC-LMS algorithm introduced in Section~\ref{subsec:ACLMS} was tested on synthetic data as an alternative to the randomized Kaczmarz's algorithm. For this experiment, $D=30,000$ observations $y_n$ were generated as in \eqref{model} with $\sigma^2=0.25$, while the $\mathbf{x}_n$'s of dimension $p=100$ were generated i.i.d. following a multivariate Gaussian distribution. For the randomized Kaczmarz's algorithm, the probability of selecting the $i-$th row is  $p_n=\|\mathbf{x}_n\|_2^2/\|\mathbf{X}\|_F^2$~\cite{strohmer2009randomized}. Since the computational complexity of the two methods is roughly the same, the comparison was done in terms of the relative MSE, namely $\mathbb{E}\Big[\|\boldsymbol{\theta}_o-\hat{\boldsymbol{\theta}}_n\|_2^2\big/{\|\boldsymbol{\theta}_o\|_2^2}\Big]$. Plotted in Fig. \ref{fig:AC_LMS}, are the relative MSE  curves of the two algorithms w.r.t. the number of data $\{\mathbf{x}_n,y_n\}$ that were used to estimate $\boldsymbol{\theta}_o$ (50 Monte Carlo runs). While the AC-LMS scans the entire dataset updating only informative data, the randomized Kaczmarz's algorithm needs access only to the data used for its updates. This is only possible if the data-dependent selection probabilities $p_n$ are given a-priori, which may not always be the case. Regardless, two more experiments were run, in which the AC-LMS had limited access to 3,000 and 1,400 data. Overall, it can be argued that when the sought reduced dimension is small, the AC-LMS offers a simple and reliable first-order alternative to the randomized Kaczmarz's algorithm.

\subsection{AC-RLS}
\label{subsec:simACRLS}
\begin{figure}[t]
	\centering
	\subfigure[]{
		\centering
		\centerline{\includegraphics[width=1\linewidth, height=2.1 in]{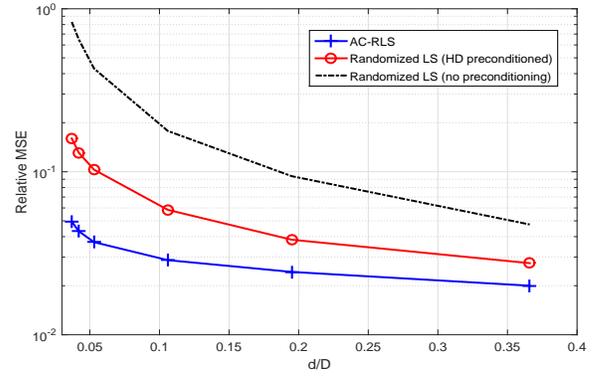}}			
		\label{fig:AC-RLs1}
	}
	\subfigure[]{
		\centering
		\centerline{\includegraphics[width=1\linewidth, height=2.1 in]{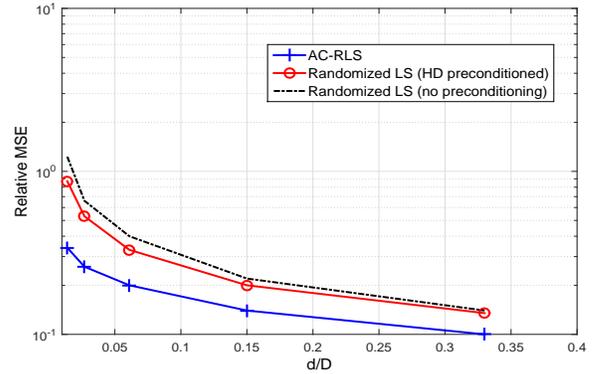}}
		\label{fig:AC-RLs2}
	}
	\subfigure[]{
	\centering
	\centerline{\includegraphics[width=1\linewidth, height=2.1 in]{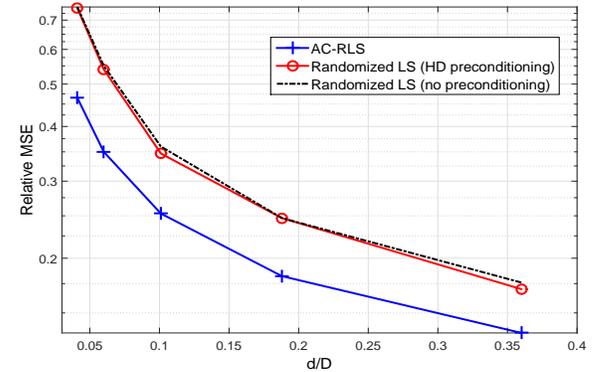}}
	\label{fig:AC-RLs3}
		}
\caption{Relative MSE of AC-RLS and randomized LS algorithms, for different levels of data reduction. Regression matrix $\textbf{X}$ was generated with highly non-uniform (a), moderately non-uniform (b), and uniform leverage scores (c).}
\end{figure}

The AC-RLS algorithm developed in Section~\ref{subsec:ACRLS} was tested on synthetic data. Specifically, the AC-RLS is treated here as an iterative method that sweeps once through the entire dataset, even though more sweeps can be performed at the cost of additional runtime. Its performance in terms of relative MSE was compared with the Hadamard (HD) preconditioned randomized LS solver, while plotted as a function of the compression ratio $d/D$. Parallel to the two methods, a uniform sampling randomized LSE was run as a simple benchmark. Measurements were generated according to \eqref{model} with $p=300$, $D=10,000$, and $v_n\sim{\mathcal{N}(0,9)}$. Regarding the data distribution, three different scenario's were examined. In Figure \ref{fig:AC-RLs1}, $\mathbf{x}_n$'s were generated according to a heavy tailed multivariate $t-$distribution with one degree of freedom, and covariance matrix with $(i,j)$-th entry $\boldsymbol{\Sigma}_{i,j}=2\times{0.5}^{|i-j|}$. Such a data distribution yields  matrices $\mathbf{X}$ with highly non-uniform leverage scores, thus imitating the effect of a subset of highly ``important'' observations randomly scattered in the dataset. In such cases, uniform sampling without preconditioning performs poorly since many of those informative measurements are missed. As seen in the plot, preconditioning significantly improves performance, by incorporating ``important'' information through random projections. Further improvement is effected by our data-driven AC-RLS through adaptively selecting the most informative measurements and ignoring the rest, without overhead in complexity.

The experiment was repeated (Fig. \ref{fig:AC-RLs2}) for $\mathbf{x}_n$ generated from a multivariate $t-$distribution with 3 degrees of freedom, and $\boldsymbol{\Sigma}$ as before. Leverage scores for this dataset are moderately non-uniform, thus inducing more redundancy and resulting in lower performance for all algorithms, while closing the ``gap'' between preconditioned and non-preconditioned random sampling. Again, the proposed AC-RLS performs significantly better in estimating the unknown parameters for the entire range of data size reduction.

Finally, Fig. \ref{fig:AC-RLs3} depicts related performance for Gaussian $\mathbf{x}_n\sim{\mathcal{N}(\mathbf{0},\boldsymbol{\Sigma})}$. Compared to the previous cases, normally distributed rows yield a highly redundant set of measurements with $\mathbf{X}$ having almost uniform leverage scores. As seen in the plots, preconditioning offers no improvement in random sampling for this type data, whereas the AC-RLS succeeds in extracting more information on the unknown $\boldsymbol{\theta}$. 

\begin{figure}[t]
\centering
\centerline{\includegraphics[width=1\linewidth, height=2.2 in]{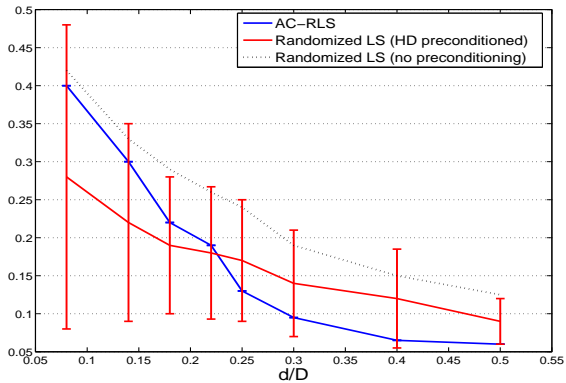}}
\caption{Relative MSE of AC-RLS and randomized LS algorithms, for different levels of data reduction using the protein tertiary structure dataset.}
\label{fig:protein}
\end{figure}

To further assess efficacy of the AC-RLS algorithm, real data tests were performed. The Protein Tertiary Structure dataset from the UCI Machine Learning Repository was tested. In this linear regression dataset, $p=9$ attributes of proteins are used to predict a value related to protein structure. A total of $D=45,730$ observations are included. Since the true $\boldsymbol{\theta}_o$ is unknown, it is estimated by solving LS on the entire dataset. Subsequently, the noise variance is also estimated via sample averaging as $\sigma^2=(1/D) \sum_{n=1}^{D}{(y_n-\mathbf{x}_n^T\boldsymbol{\theta}_o)^2}$. Figure \ref{fig:protein} depicts relative squared-error (RSE) with respect to the data reduction ratio $d/D$. The RSE curve for the HD-preconditioned LS corresponds to the average RSE across 50 runs, while the size of the vertical bars is proportional to its standard deviation. Different from RP-based methods, the RSE for AC-RLS does not entail standard deviation bars, because for a given initialization and data order, the output of the algorithm is deterministic. It can be observed that for $d/D\geq{0.25}$ the AC-RLS outperforms RPs in terms of estimating $\boldsymbol{\theta}$, while 
for very small $d/D$, RPs yield a lower average RSE, at the cost however of very high error uncertainty (variance).    

\subsection{Robust AC-RLS}
\label{subsec:simrobustACRLS}
\begin{figure}[t]
	\centering
	\centerline{\includegraphics[width=1\linewidth, height=2.1 in]{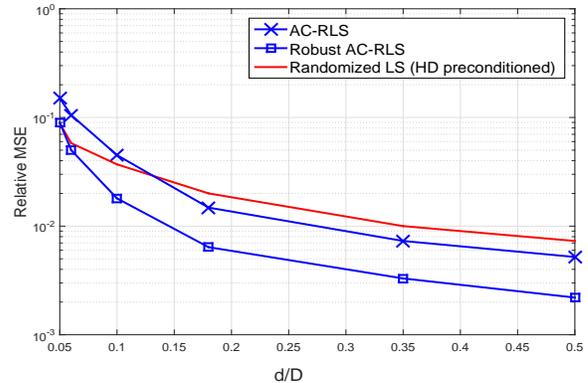}}
	\caption{Relative MSE of AC-RLS, rAC-RLS, and randomized LS algorithms, for different levels of data reduction using an outlier-corrupted dataset.}
	\label{fig:robustAC-RLS}
\end{figure}

To test rAC-LMS and rAC-RLS of Section~\ref{subsec:robust}, datasets were generated with $D=10,000$, $p=30$ and $\mathbf{x}_n\sim{\mathcal{N}(\mathbf{0},\boldsymbol{\Sigma})}$, where $\boldsymbol{\Sigma}_{i,j}=2\times{0.5}^{|i-j|}$; noise was i.i.d. Gaussian $v_n\sim{\mathcal{N}(0,9)}$; meanwhile measurements $y_n$ were generated according to \eqref{model} with random and sporadic outlier spikes $\{o_n\}_{n=1}^D$. Specifically, we generated $o_n=\alpha_n\beta_n$, where $\alpha_n\sim{\textrm{Bernoulli}}(0.05)$, and $\beta_n\sim{\mathcal{N}(0,25\times{9})}$, thus resulting in approximately $5\%$ of the data effectively being outliers. Similar to previous experiments, our novel algorithms were run once through the set selecting $d$ out of $D$ data to update $\boldsymbol{\theta}_n$. Plotted in Fig. \ref{fig:robustAC-RLS} is the RSE averaged across 100 runs as a function of $d/D$ for the HD-preconditioned LS, the plain AC-RLS, and the rAC-RLS with a Huber-like instantaneous cost. As expected, the performance of AC-RLS is severely undermined especially when tuned for very small $d/D$, exhibiting higher error than the RP-based LS. However, our rAC-RLS algorithm offers superior performance across the entire range of $d/D$ values.     

\section{Concluding Remarks}\label{sec:conclusion}
We developed online algorithms for large-scale LS linear regressions that rely on censoring for data-driven dimensionality reduction of streaming Big Data. First, a non-adaptive censoring setting was considered for applications where observations are censored -- possibly naturally -- separately and prior to estimation. Computationally efficient first- and second-order online algorithms were derived to estimate the unknown parameters, relying on stochastic approximation of the log-likelihood of the censored data. Performance was bounded analytically, while simulations demonstrated that the second-order method performs close to the CRLB.

Furthermore, online data reduction occurring parallel to estimation was also explored. For this scenario, censoring is performed deliberately and adaptively based on estimates provided by first- and second-order algorithms. Robust versions were also developed for estimation in the presence of outliers. Studied under the scope of stochastic approximation, the proposed algorithms were shown to enjoy guaranteed MSE performance. Moreover, the resulting recursive methods were advocated as low-complexity recursive solvers of large LS problems. Experiments run on synthetic and real datasets corroborated that the novel AC-LMS and AC-RLS algorithms outperformed competing randomized algorithms.

Our future research agenda includes approaches to nonlinear (e.g., kernel-based) parametric and nonparametric large-scale regressions, along with estimation of dynamical (e.g., state-space) processes using adaptively censored measurements.

\appendix
\begin{IEEEproof}[Proof of Proposition~\ref{pro:regret}]
It can be verified that $\nabla^2\ell_n(\boldsymbol{\theta})\succeq{\boldsymbol{0}}$, which implies the convexity of $\ell_n(\boldsymbol{\theta})$~\cite{tsp2012ggeric}. The regret of the SGD approach is then bounded as~\cite[Corollary 2.7]{shalev2011online}
\begin{align*}
R(D) & \le\frac{1}{2\mu}\|\boldsymbol{\theta}^\ast-\boldsymbol{\theta}_1\|_2^2+\mu\sum_{n=1}^D\|\nabla \ell_n(\boldsymbol{\theta}_{n-1})\|_2^2\\ 
& = \frac{1}{2\mu} \|\boldsymbol{\theta}^\ast-\hat{\boldsymbol{\theta}}_K\|_2^2+\mu\sum_{n=1}^D\|\mathbf{x}_n\|_2^2\beta^2(\boldsymbol{\theta}_{n-1})\\
&\le\frac{1}{2\mu}\|\boldsymbol{\theta}^\ast-\hat{\boldsymbol{\theta}}_K\|_2^2+\mu D (\bar{x}\bar{\beta})^2
\end{align*}
where $\{\boldsymbol{\theta}_n\}_{n=1}^D$ is any sequence of estimates produced by the SA-MLE. By choosing $\mu=\|\boldsymbol{\theta}^\ast-\hat{\boldsymbol{\theta}}_K\|_2 /(\sqrt{2D}\bar\beta\bar{x})$, the aforementioned bound leads to Proposition~\ref{pro:regret}.
\end{IEEEproof}

\begin{IEEEproof}[Proof of Proposition~\ref{pro:converge_aclms}] For the SGD update in~\eqref{eq:ACLMS}, the MSE $\mathbb{E}_{\mathbf{x},v}\left[{\|\boldsymbol{\theta}_n-\boldsymbol{\theta}_o\|_2^2}\right]$, with $\boldsymbol{\theta}_o=\arg\min_{\boldsymbol{\theta}} F(\boldsymbol{\theta})$ where $F(\boldsymbol{\theta}):=\mathbb{E}_{\mathbf{x},v}\left[f^{(\tau)}(\boldsymbol{\theta};\mathbf{y})\right]$ is bounded as in~\cite{bach}. For this to hold, we must have: a1) the gradient bounded at the optimum; that is,  $\mathbb{E}_{\mathbf{x},v}\left[\|\nabla{f^{(\tau)}(\boldsymbol{\theta}_o,\mathbf{y})}\|_2^2\right]\leq{\Delta}$; a2) the gradient must be $L-$smooth for any other $\boldsymbol{\theta}$; and a3) $F(\boldsymbol{\theta})$ must be $\alpha$-strongly convex~\cite{bach}. With $\mathbf{x}$ and $v$ generated randomly and independently across time, associated quantities do not depend on $n$. Furthermore, the points of discontinuity of $f^{(\tau)}(.)$ are zero-measure in expectation, and thus are neglected for brevity.  

Under a3), there exists a constant $\alpha>0$ such that $\nabla^2F(\boldsymbol{\theta})\succeq{\alpha\mathbf{I}}$ $\forall \boldsymbol{\theta}$. Interchanging differentiation with expectation yields
\begin{align*}
         \nabla^2F(\boldsymbol{\theta})&=\nabla^2_{\boldsymbol{\theta}}\mathbb{E}_{\mathbf{x},v}\left[f^{(\tau)}(\boldsymbol{\theta};\mathbf{x},v)\right]\\
         &=
         \mathbb{E}_{\mathbf{x},v}\left[\nabla^2_{\boldsymbol{\theta}}\frac{e^2}{2}(1-c)\right]=\mathbb{E}_{\mathbf{x},v}\left[\mathbf{x}\mathbf{x}^T(1-c)\right]\\
         &=
         \int_{\mathbf{x}}\int_{v}\mathbf{x}\mathbf{x}^T\mathbbm{1}_{\{|\mathbf{x}^T(\boldsymbol{\theta}_o-\boldsymbol{\theta})+v|\geq{\tau\sigma}\}}p_v(v)p_x(\mathbf{x})d{v}d{\mathbf{x}}\\
         &=
         \int_{\mathbf{x}}\mathbf{x}\mathbf{x}^T\left(\int_{v}\mathbbm{1}_{\{|\mathbf{x}^T(\boldsymbol{\theta}_o-\boldsymbol{\theta})+v|\geq{\tau\sigma}\}}p_v(v)d{v}\right)p_x(\mathbf{x})d{\mathbf{x}}\\
         &=
         \int_{\mathbf{x}}\mathbf{x}\mathbf{x}^T\bigg[1-Q\left(-\tau-\frac{\mathbf{x}^T(\boldsymbol{\theta}_o-\boldsymbol{\theta})}{\sigma}\right)\\
         &+Q\left(\tau-\frac{\mathbf{x}^T(\boldsymbol{\theta}_o-\boldsymbol{\theta})}{\sigma}\right)\bigg]p_x(\mathbf{x})d{\mathbf{x}}\\   
         &=
         \int_{\mathbf{x}}\mathbf{x}\mathbf{x}^T\bigg[Q\left(\tau+\frac{\mathbf{x}^T(\boldsymbol{\theta}_o-\boldsymbol{\theta})}{\sigma}\right)\\
         &+Q\left(\tau-\frac{\mathbf{x}^T(\boldsymbol{\theta}_o-\boldsymbol{\theta})}{\sigma}\right)\bigg]p_x(\mathbf{x})d{\mathbf{x}}.               
      \end{align*}
It can be verified that the function $g(z):=Q(\tau+z)+Q(\tau-z)$ is minimized for $z=0$ when $\tau>0$. To see this, observe that its derivative $g'(z)=-\phi(\tau+z)+\phi(\tau-z)$ vanishes when $|\tau+z|=|\tau-z|$. Therefore, $g(z)\geq g(0)=2Q(\tau)$ for all $z$; and hence,
\begin{equation*}
      Q\left(\tau+\frac{\mathbf{x}^T(\boldsymbol{\theta}_o-\boldsymbol{\theta})}{\sigma}\right)+Q\left(\tau-\frac{\mathbf{x}^T(\boldsymbol{\theta}_o-\boldsymbol{\theta})}{\sigma}\right)\geq{2Q(\tau)}
\end{equation*}
for all $\mathbf{x}$ and $\boldsymbol{\theta}$. The latter implies
\begin{align*}
\nabla^2F(\boldsymbol{\theta}) & \succeq \int_{\mathbf{x}} \mathbf{x}\mathbf{x}^T 2Q(\tau) p_x(\mathbf{x})d{\mathbf{x}} =2Q(\tau)\mathbf{R}_{x}\\
&\succeq 2Q(\tau)\lambda_{\min}(\mathbf{R}_{x})\mathbf{I}
\end{align*} 
showing that $F(\boldsymbol{\theta})$ is $\alpha-$strongly convex with $\alpha =2Q(\tau)\lambda_{\min}(\mathbf{R}_{x})$. As expected, $\alpha$ reduces for increasing $\tau$. 
   
Regarding the instantaneous gradient, it suffices to find $L$ such that
$ \mathbb{E}_{\mathbf{x},v}\left[\|\nabla{f^{(\tau)}(\boldsymbol{\theta}_1)}-\nabla{f^{(\tau)}(\boldsymbol{\theta}_2)}\|_2^2\right]\leq{L^2\|\boldsymbol{\theta}_1-\boldsymbol{\theta}_2\|_2^2}$ for all $n$ and any pair $(\boldsymbol{\theta}_1,\boldsymbol{\theta}_2)$. For the errors $\boldsymbol{\zeta}_i:=\boldsymbol{\theta}_o-\boldsymbol{\theta}_i$ for $i=1,2$, it holds
\begin{align}\nonumber
   &\mathbb{E}_{\mathbf{x},v}\left[\|\nabla{f^{(\tau)}(\boldsymbol{\theta}_1)}-\nabla{f^{(\tau)}(\boldsymbol{\theta}_2)}\|_2^2\right]\\\nonumber
   &=
   \mathbb{E}_{\mathbf{x},v}\left[{\|\mathbf{x}e(\boldsymbol{\theta}_1)(1-c_1)-\mathbf{x}e(\boldsymbol{\theta}_2)(1-c_2)}\|_2^2\right]\\\nonumber
   &=
   \mathbb{E}_{\mathbf{x},v}\bigg[\|\mathbf{x}(\mathbf{x}^T\boldsymbol{\zeta}_1+v)\mathbbm{1}_{\{|\mathbf{x}^T\boldsymbol{\zeta}_1+v|\geq{\tau\sigma}\}}\\\nonumber
   	&-\mathbf{x}(\mathbf{x}^T\boldsymbol{\zeta}_2+v)\mathbbm{1}_{\{|\mathbf{x}^T\boldsymbol{\zeta}_2+v|\geq{\tau\sigma}\}}\|_2^2\bigg]\\\nonumber
   &=
    \mathbb{E}_{\mathbf{x},v}\big[\|\mathbf{x}\mathbf{x}^T\boldsymbol{\zeta}_1\mathbbm{1}_{\{|\mathbf{x}^T\boldsymbol{\zeta}_1+v|\geq{\tau\sigma}\}}-\mathbf{x}\mathbf{x}^T\boldsymbol{\zeta}_2\mathbbm{1}_{\{|\mathbf{x}^T\boldsymbol{\zeta}_2+v|\geq{\tau\sigma}\}}\\\nonumber &+\mathbf{x}v(\mathbbm{1}_{\{|\mathbf{x}^T\boldsymbol{\zeta}_1+v|\geq{\tau\sigma}\}}-\mathbbm{1}_{\{|\mathbf{x}^T\boldsymbol{\zeta}_2+v|\geq{\tau\sigma}\}})\|_2^2\big]\\\nonumber
    &=
    \mathbb{E}_{\mathbf{x},v}\bigg[\boldsymbol{\zeta}_1^T\left(\mathbf{x}\mathbf{x}^T\right)^2\mathbbm{1}_{\{|\mathbf{x}^T\boldsymbol{\zeta}_1+v|\geq{\tau\sigma}\}}+\boldsymbol{\zeta}_2^T\left(\mathbf{x}\mathbf{x}^T\right)^2\mathbbm{1}_{\{|\mathbf{x}^T\boldsymbol{\zeta}_2+v|\geq{\tau\sigma}\}}\\ \nonumber
    &-2\boldsymbol{\zeta}_1^T\left(\mathbf{x}\mathbf{x}^T\right)^2\boldsymbol{\zeta}_2\mathbbm{1}_{\{|\mathbf{x}^T\boldsymbol{\zeta}_1+v|\geq{\tau\sigma}\}}\mathbbm{1}_{\{|\mathbf{x}^T\boldsymbol{\zeta}_2+v|\geq{\tau\sigma}\}}\\\nonumber
    &+\mathbf{x}^T\mathbf{x}\mathbf{x}^T\boldsymbol{\zeta}_1\mathbbm{1}_{\{|\mathbf{x}^T\boldsymbol{\zeta}_1+v|\geq{\tau\sigma}\}}v\left(\mathbbm{1}_{\{|\mathbf{x}^T\boldsymbol{\zeta}_1+v|\geq{\tau\sigma}\}}-\mathbbm{1}_{\{|\mathbf{x}^T\boldsymbol{\zeta}_2+v|\geq{\tau\sigma}\}}\right)\\\nonumber
    &-\mathbf{x}^T\mathbf{x}\mathbf{x}^T\boldsymbol{\zeta}_2\mathbbm{1}_{\{|\mathbf{x}^T\boldsymbol{\zeta}_2+v|\geq{\tau\sigma}\}}v\left(\mathbbm{1}_{\{|\mathbf{x}^T\boldsymbol{\zeta}_1+v|\geq{\tau\sigma}\}}-\mathbbm{1}_{\{|\mathbf{x}^T\boldsymbol{\zeta}_2+v|\geq{\tau\sigma}\}}\right)\\
    &+\|\mathbf{x}\|_2^2v^2\left(\mathbbm{1}_{\{|\mathbf{x}^T\boldsymbol{\zeta}_1+v|\geq{\tau\sigma}\}}-\mathbbm{1}_{\{|\mathbf{x}^T\boldsymbol{\zeta}_2+v|\geq{\tau\sigma}\}}\right)^2\bigg]\label{huge}.
   \end{align}
   It can be verified that since the cross-terms in~\eqref{huge} can be bounded from below and above as 
\begin{align*}
&\mathbb{E}_{\mathbf{x}}\left[\mathbf{x}^T\mathbf{x}\mathbf{x}^T\right]\boldsymbol{\zeta}_1L(\boldsymbol{\zeta}_1,\boldsymbol{\zeta}_2)\leq\mathbb{E}_{\mathbf{x}}\big[\mathbf{x}^T\mathbf{x}\mathbf{x}^T\boldsymbol{\zeta}_1\\
&\times\mathbb{E}_v\left[\mathbbm{1}_{\{|\mathbf{x}^T\boldsymbol{\zeta}_1+v|\geq{\tau\sigma}\}}v\left(\mathbbm{1}_{\{|\mathbf{x}^T\boldsymbol{\zeta}_1+v|\geq{\tau\sigma}\}}-\mathbbm{1}_{\{|\mathbf{x}^T\boldsymbol{\zeta}_2+v|\geq{\tau\sigma}\}}\right)\right]\big]\\
&\leq
\mathbb{E}_{\mathbf{x}}\left[\mathbf{x}^T\mathbf{x}\mathbf{x}^T\right]\boldsymbol{\zeta}_1U(\boldsymbol{\zeta}_1,\boldsymbol{\zeta}_2),
\end{align*}
 they are also equal to zero if the third-order moment $\mathbb{E}_{\mathbf{x}}\left[\mathbf{x}^T\mathbf{x}\mathbf{x}^T\right]=\mathbf{0}$. Furthermore, by simply bounding $\mathbb{E}_v\left[\mathbbm{1}_{\{|\mathbf{x}^T\boldsymbol{\zeta}_i+v|\geq{\tau\sigma}\}}\right]\leq{1}$ as probabilities, \eqref{huge} yields
\begin{align*}
&\mathbb{E}\left[\|\nabla{f(\boldsymbol{\theta}_1)}-\nabla{f(\boldsymbol{\theta}_2)}\|_2^2\right]\leq\mathbb{E}_{\mathbf{x}}\bigg[(\boldsymbol{\zeta}_1-\boldsymbol{\zeta}_2)^T\left(\mathbf{x}\mathbf{x}^T\right)^2(\boldsymbol{\zeta}_1-\boldsymbol{\zeta}_2)\\
&+\|\mathbf{x}\|_2^2\mathbb{E}_v\left[v^2\left(\mathbbm{1}_{\{|\mathbf{x}^T\boldsymbol{\zeta}_1+v|\geq{\tau\sigma}\}}-\mathbbm{1}_{\{|\mathbf{x}^T\boldsymbol{\zeta}_2+v|\geq{\tau\sigma}\}}\right)^2\right]\bigg]\\
&=
(\boldsymbol{\zeta}_1-\boldsymbol{\zeta}_2)^T\mathbb{E}_{\mathbf{x}}\left[\left(\mathbf{x}\mathbf{x}^T\right)^2\right](\boldsymbol{\zeta}_1-\boldsymbol{\zeta}_2)\\
&+\mathbb{E}_{\mathbf{x}}\left[\|\mathbf{x}\|_2^2\mathbb{E}_v\left[v^2\left(\mathbbm{1}_{\{|\mathbf{x}^T\boldsymbol{\zeta}_1+v|\geq{\tau\sigma}\}}-\mathbbm{1}_{\{|\mathbf{x}^T\boldsymbol{\zeta}_2+v|\geq{\tau\sigma}\}}\right)^2\right]\right]\\
&\leq
\left(\lambda_{\max}\left(\mathbb{E}\left[\left(\mathbf{x}\mathbf{x}^T\right)^2\right]\right)+\lambda_{\tau}\right)\|\boldsymbol{\theta}_1-\boldsymbol{\theta}_2\|_2^2.
\end{align*}
The last expression reveals that the average distance between gradients can be decomposed into two terms. The first term can be bounded using the fourth-order moment. The second term appears due to data censoring and clearly depends on $\tau$, while it is assumed bounded as
\begin{align*}
&\mathbb{E}_{\mathbf{x}}\left[\|\mathbf{x}\|_2^2\mathbb{E}_v\left[v^2\left(\mathbbm{1}_{\{|\mathbf{x}^T\boldsymbol{\zeta}_1+v|\geq{\tau\sigma}\}}-\mathbbm{1}_{\{|\mathbf{x}^T\boldsymbol{\zeta}_2+v|\geq{\tau\sigma}\}}\right)^2\right]\right]\\
&\leq{\lambda_{\tau}\|\boldsymbol{\theta}_1-\boldsymbol{\theta}_2\|_2^2}.
\end{align*}
Although we could not express $\lambda_{\tau}$ in closed form, for relatively small values of $\tau$ used in practice to censor more than $90\%$ of the measurements, $\lambda_{\tau}\approx{0}$; thus, the second term can be ignored yielding $L^2\approx\lambda_{\max}\left(\mathbb{E}\left[\left(\mathbf{x}\mathbf{x}^T\right)^2\right]\right)$. Furthermore, even for large $\tau$ some inaccuracy in the value of $L$ can be tolerated, after considering that it does not affect the algorithm's stability or asymptotic performance when a vanishing step size is used.

Finally, the expected norm of the gradient at $\boldsymbol{\theta}=\boldsymbol{\theta}_o$ is bounded and equal to
\begin{align*}
&\mathbb{E}\left[\|\nabla f^{(\tau)}(\boldsymbol{\theta}_o)\|_2^2\right]=\mathbb{E}\left[\|\mathbf{x}\|_2^2e(\boldsymbol{\theta}_o)(1-c)\right]\\
&=\mathbb{E}_{\mathbf{x}}\left[\|\mathbf{x}\|_2^2\right]\mathbb{E}_v\left[v^2\mathbbm{1}_{\{|v|>\tau\sigma\}}\right]\\
&=\mathrm{tr}(\mathbf{R}_{x})\left[\sigma^2-\int\limits_{-\tau\sigma}^{\tau\sigma}v^2\frac{e^{-\frac{v^2}{2\sigma^2}}}{\sqrt{2\pi\sigma^2}}dv\right]\\
&=\mathrm{tr}(\mathbf{R}_{x})\left[\sigma^2-\sigma^2\left[Q\left(\frac{v}{\sigma}\right)-\frac{v}{\sigma}\phi\left(\frac{v}{\sigma}\right)\right]_{-\tau\sigma}^{\tau\sigma}\right]\\
& = 2\sigma^2\mathrm{tr}(\mathbf{R}_{x})\left(1-Q(\tau)+\tau \phi(\tau)\right) 
\end{align*}  
which completes the proof.
\end{IEEEproof}

\begin{IEEEproof}[Proof of Proposition~\ref{pro:converge_acrls}] 
For the error vector $\boldsymbol{\zeta}_n := \boldsymbol{\theta}_n-\boldsymbol{\theta}_o$, AC-RLS satisfies $\boldsymbol{\zeta}_n=\mathbf{C}_n\sum\limits_{i=1}^{n}\mathbf{x}_iv_i(1-c_i)$. If $\{c_i\}_{i=1}^{n}$ are deterministic and given, the error covariance matrix 
$\mathbf{K}_n :=\mathbb{E} [\boldsymbol{\zeta}_n\boldsymbol{\zeta}_n^T]$ becomes
\begin{align*}
\mathbf{K}_n&=\mathbb{E}_{\mathbf{x},v} \left[\mathbf{C}_n\sum\limits_{i=1}^{n}\sum\limits_{j=1}^{n}\mathbf{x}_i\mathbf{x}_j^Tv_iv_j(1-c_i)(1-c_j)\mathbf{C}_n \right]\\
&=\mathbb{E}_{\mathbf{x}} \left[\mathbf{C}_n\sum\limits_{i=1}^{n}\sum\limits_{j=1}^{n}\mathbf{x}_i\mathbf{x}_j^T\mathbb{E}_v\left[v_iv_j\right](1-c_i)(1-c_j)\mathbf{C}_n \right]\\
&=
\sigma^2\mathbb{E}_{\mathbf{x}} \left[\mathbf{C}_n\sum\limits_{i=1}^{n}\mathbf{x}_i\mathbf{x}_i^T(1-c_i)\mathbf{C}_n \right]\\
&=
\sigma^2\mathbb{E}_{\mathbf{x}}\left[\mathbf{C}_n\mathbf{C}_n^{-1}\mathbf{C}_n\right]=\sigma^2\mathbb{E}_{\mathbf{x}} [\mathbf{C}_n ]
\end{align*}  
Assuming $\mathbf{x}_n\mathbf{x}_n^T(1-c_n)$ to be ergodic and for large enough $n$, the matrix $\mathbf{C}_n^{-1}=\sum\limits_{i=1}^{n}\mathbf{x}_i\mathbf{x}_i^T(1-c_i)$ can be approximated by $n\mathbb{E}_{\mathbf{x},v}\left[\mathbf{x}\mathbf{x}^T(1-c)\right]=n\mathbb{E}_{\mathbf{x}}\left[\mathbf{x}\mathbf{x}^T\mathbb{E}_v[1-c]\right]=n\mathbb{E}_{\mathbf{x}}\left[\mathbf{x}\mathbf{x}^T\Pr\{c=0|\mathbf{x}\}\right]=\mathbf{C}_{\infty}^{-1}$. Given that $2Q(\tau)\leq{\Pr\{c=0|\mathbf{x}\}}\leq{1}\:\forall\mathbf{x}$, we obtain \[2Q(\tau)n\mathbf{R}_{x}\preceq{\mathbf{C}_{\infty}^{-1}}\preceq{n\mathbf{R}_{x}}.\]
Since $\mathbf{C}_n$ converges monotonically to $\mathbf{C}_{\infty}$, there exists $k>0$ such that for all $n>k$
 \begin{equation*}
\frac{1}{n}\mathbf{R}_{x}^{-1}\preceq{\mathbf{C}_{n}}\preceq \frac{1}{2Q(\tau)n} \mathbf{R}_{x}^{-1}.
 \end{equation*}
The result follows given that $\mathbb{E}\left[\|\boldsymbol{\theta}_n-\boldsymbol{\theta}_o\|_2^2\right]=\mathrm{tr}(\mathbf{K}_n)=\sigma^2\mathrm{tr}(\mathbb{E}\left[\mathbf{C}_n\right])$.
\end{IEEEproof}
\bibliographystyle{IEEEtran}
\bibliography{ref_censoring}
\end{document}